\documentclass[a4paper,reqno,twoside,psamsfonts]{amsart}
\usepackage[latin1]{inputenc}
\usepackage[T1]{fontenc}
\usepackage{textcomp}
\usepackage[english]{babel}
\usepackage{pstricks,pst-plot,pst-grad,pst-node,graphicx,ifthen}
\usepackage{newcent} 
\usepackage{AUS-math}
\providecommand{\href}[2]{#2}

\providecommand{\hypersetup}[1]{}
\newif\ifpdf\ifx\pdfoutput\undefined\pdffalse\else\pdftrue\fi
\ifpdf 
  \hypersetup{%
    pdftitle= {A Model for Prejudiced Learning in Noisy Environments},
    pdfauthor= {Andreas U. Schmidt},
    pdfsubject= {2000 Mathematical subject classification: 91A26, 34F05},
    pdfkeywords= { learning, prejudice, uncertainty, noise, random dynamical system,
noise induced stability, stochastic bifurcation } } 
\else\fi 
\newcommand{\PD}[2][\rho,\sigma]{%
\ifthenelse{\equal{#2}{}}%
{\ensuremath{\mathord{{\mathcal{P}}_{#1}}}}%
{\ensuremath{\mathord{{\mathcal{P}}_{#1}(#2)}}}} 
\begin{document}\selectlanguage{english}
%
%
%
%
\title[Prejudiced Learning in Noisy Environments]{A Model for Prejudiced Learning in Noisy Environments}
\author[A.\ U.\ Schmidt]{Andreas U.\ Schmidt}
\date{6th August 2004}
\address{Fraunhofer -- Institute for Secure Information Technology\\
Dolivostraße 15\\
64293 Darmstadt\\
Germany}
\email{\href{mailto:aschmidt@math.uni-frankfurt.de}{aschmidt@math.uni-frankfurt.de}}
\urladdr{\href{http://www.math.uni-frankfurt.de/~aschmidt}{http://www.math.uni-frankfurt.de/\~{}aschmidt}}
\thanks{This research was partially supported by a project within the fifth
framework programme's priority `Information Society Technologies' of the EU.
The provision of parallel computing resources by the GMD institute SCAI 
(now Fraunhofer SCAI) is gratefully acknowledged.}
\begin{abstract}
 Based on the heuristics that maintaining presumptions can be 
beneficial in uncertain environments, we propose a set of basic 
requirements for learning systems to incorporate the concept of prejudice.
The simplest, memoryless model of a deterministic learning rule obeying
the axioms is constructed, and shown to be equivalent to the logistic map. 
The system's performance is analysed
in an environment in which it is subject to external randomness, 
weighing learning defectiveness against stability gained.
 The corresponding random dynamical system with inhomogeneous, additive noise
is studied, and shown to exhibit the phenomena of noise 
induced stability and stochastic bifurcations. 
The overall results allow for the interpretation that prejudice 
in uncertain environments can entail a 
considerable portion of stubbornness as a secondary phenomenon.
\end{abstract}
\maketitle
%
%
\section{Introduction}
As almost all terms denoting affects,
the term `prejudice' is as ubiquitous as ill-defined, 
and resists na\"ive attempts to provide it with meaning 
in any more  epistemologically rigorous sense. 
Yet the last century saw, with the advent of game theory
as a means of formalisation and modelling,
the paving of a scientific access path 
to such notions, partially as a side effect of the growing
interest in the behaviour of intelligent beings
(usually called agents) in social environments~\cite{TESF}, 
which in turn was spurred by the mathematisation of economics.
With the emergence of the Internet
as a mass medium, this kind of research 
has obtained a new test bed, a source of 
statistical data, and an independent study object,
and has thus gained further impetus~\cite{TF99,TF99A,GO01}.
Consequently, hard science is lead to occupy itself
with formerly foreign concepts from the 
psychological and sociological domains. 
Recent efforts in this direction
combine game theory and logic with nonlinear systems
theory and stochastics, and some of them are quite bold, 
as for instance~\cite{RIN98}.
The present study parallels these lines of thought, and is,
to the best of our knowledge, the first one to 
focus on the concept of prejudice.
It is based on the report~\cite{PREDI}, but is
essentially enlarged and refined.

\refS{back} provides the theoretical framework on which
the subsequent construction of a  model for prejudice rests.
We set up a simple, general model for an environment
in which an agent needs to learn a fluctuating \emph{risk}, and
argue that prejudice might be beneficial in such circumstances.

The special instance of a prejudiced learning rule we consider
is constructed in~\refS{Model}.
In particular, we define what we mean by a \emph{prejudiced}
deterministic learning rule, in the language of reinforcement 
learning.
We then set out three 
axioms we regard as heuristically sound for prejudiced 
learning in environments polluted with noise.
We then look for the most simple prejudiced learning 
rule fulfilling them, 
a rule which in particular does not require the agents to have memory. 
In the noiseless case, the ensuing rule turns out 
to be equivalent to the logistic map. In correspondence 
to the well known dynamical features of this map
we classify the general behaviour of prejudiced learners,
and obtain restrictions on their internal parameters.
When the input to the prejudiced learning rule is subject to
random fluctuations, it still retains the form of the logistic
map, but with additively coupled, inhomogeneous noise.

To corroborate the argument that the limitation of rationality
presented by prejudiced learning can be beneficial in certain circumstances,
a numerical analysis of the performance of a subclass 
of prejudiced learners in an uncertain environment
is carried out in \refS{perf}, for various levels of noise. 
The result is twofold. On the one hand, the prejudice causes the agents 
to make a small error in their belief about the true risk, 
generally overestimating it. On the other hand, prejudiced learning
can efficiently stabilise an agent's behaviour, in particular
for higher values of the logistic map's single dynamical parameter. 
Therefore, if both factors are taken into account,
prejudiced learning has the potential to be advantageous 
in noisy environments.

Our special prejudiced learning rule with noise is an example for
a \emph{noisy dynamical system}, a class of systems which has attracted
a lot of interest from physicists and mathematicians in recent 
years. Reference~\cite{CP83} is one earlier, seminal work.
Noisy, or random dynamical systems show a host of additional phenomenology
over ordinary ones, including phenomena that are to be expected
from natural systems. Furthermore, they present a combination
of nonlinearity and stochastics on which now a whole branch of mathematics
thrives, see~\cite{b:ARN98} and its vast bibliography.
The system defined by the prejudiced learning rule
is therefore interesting in its own right, and we devote \refS{NDS}
to its study. In it, we rediscover the phenomenon of
\textit{stochastic resonance} or
\textit{noise induced stability}~\cite{BSV81,BPSV83}, for a large range of parameter
values and noise levels, as well as the phenomenon of \textit{stochastic
bifurcations} or \textit{noise induced transitions}~\cite{FMMM87,NA87} 
as these parameters vary. Both phenomena have been empirically 
confirmed in a vast variety of natural systems and models thereof,
ranging from biophysics~\cite{YHTI02} 
and chemistry~\cite{DHR95}, over financial markets~\cite{KH03} 
and signal processing~\cite{CBG97}, to quantum information
theory~\cite{LJ02A}, without any claim to completeness.
We determine the stability domain  of our prejudiced learning
system analytically using the Lyapunov exponent, and study the bifurcating
transition from that domain, which is evoked by lowering the noise level, using
numerics. In particular, we determine the \textit{critical exponent} 
of this transition, in analogy to concepts of statistical mechanics.

Finally, \refS{Conc} contains a comprising assessment of our
model for prejudiced learning. It concludes with some suggestions
for further research to be based on this and similar models.
\section{Background and Heuristics}
\labelS{back}
To place our work into theoretical context and delineate its scope 
we briefly recapitulate some background.
The framework for the construction of our model for prejudiced learning
is that of basic game theory and reinforcement learning~\cite{b:LR89,b:SB98}. 
We briefly sketch the necessary background.

The classical, game-theoretical subject of 
\emph{decision making under uncertainty} 
considers single- or multi player games of 
intelligent agents against an environment (nature). 
Each agent has an \emph{utility function} depending 
on the action he takes and the (unknown) state of nature. 
This function describes the payoff of the execution of the 
corresponding action given that nature is in the corresponding state. 
The utility function is taken to be the input for the agent's 
\emph{decision rule} which uniquely determines the action to be taken. 
In the more realistic case when a probability distribution over the states 
of nature is known to the agent, one is in the realm of 
\emph{statistical decision theory}. 
There, this knowledge is fed into a selection scheme 
which determines one from a set of 
decision rules accordingly. For example, the well 
known \emph{Bayesian decision rule} 
is the combination of a selection scheme and a decision 
rule which assigns to each action the 
average sum of utilities weighted with the known 
probability distribution, and then chooses 
the action maximising this value. The knowledge can 
be \emph{a priori} or learnt by 
experimentation using some 
\emph{statistical learning rule}, which can be as simple as 
taking means over a finite set of experimental 
outcomes (\emph{Bayesian learning}). 
When the result of taking a specific action is 
fed back as the outcome of an experiment 
into the learning rule and the whole cycle is 
repeated many times, the agent becomes a 
\emph{learning automaton}. This is the basic object we consider.

The Bayesian rule is obviously not the only possible, 
there is a multitude of 
different learning rules and selection schemes. 
Numerous statistical learning rules have been considered in pursuit of
the optimal one in a given environment, 
see\eg~\cite{b:VAP00,b:HTF01}, 
and references therein.
In realistic cases it can be sensible 
to choose selection schemes which differ significantly from the apparently most 
rational Bayesian rule. Here, we propose a model which can justifiably be said to 
represent agents who are \emph{prejudiced} by construction and in behaviour. 
We derive the heuristics for our construction from a key example:
\begin{xmpl}\label{xmpl:sec}
At each time step, the agent takes an action, selected from a finite set $A$,
 with a certain payoff 
whose maximal value we normalise to $1$, for simplicity.  
Choosing action $k$, there is a --- generically small --- probability $p_k$ 
for the occurrence of a damage $d_k$ which diminishes the maximal payoff to $1-d_k$. 
Assume that the actual damage is symmetrically distributed with 
small variance around the mean 
value $\overline{d}_k\in[0,1]$. 
The expected payoff then is the (true empirical) \textbf{weight} 
$\overline{w}_k\equiv 1-\overline{r}_k\equiv 1-\overline{d}_k\cdot p_k$ 
of action $k$ and $\overline{r}_k$ is its true \textbf{risk}.
The weights are the quantities rational decision makers would
base their decisions on.
Approximations for them are learnt by the agents from the frequency and 
amount of previous damages they actually incurred. 
These are fed into a selection scheme which in turn 
determines the decision rule for action selection.
\end{xmpl}
The general feature of this example rendering the application of Bayes' rule less 
attractive is that the damage rate $1/p_k$ and therefore the learning speed of any 
statistical inference rule for the set $\{\overline{r}_k\}$ 
can be very low compared to the frequency 
with which an action has to be taken. 
Thus, initial probabilistic fluctuations could result in the (costly) selection 
of a non-optimal action for many steps. On the other hand, the expected damage 
$d_k$ can be close to $1$, resulting in relatively high risk. 
Thus, the agent has high 
interest in using a reliable \emph{a priori} risk estimate, to 
keep it stable against fluctuations, and still learn as 
quickly as possible.

Numerous studies in algorithmic learning theory are concerned
with optimal learning performance in adverse environments polluted
with noise, see~\cite{BEK02A} and references therein.
There, the strategies of the learner, and of its adversaries creating 
the noise, are rather elaborate. 
Our present approach differs from these by focusing on constructive
simplicity of the model and its behavioural features rather than
optimality. Furthermore, our adversary will be a single, 
very simple, noise model.
\section{Model Building}
\labelS{Model}
\begin{figure}[tb]
\psset{xunit=8mm,yunit=8mm}
\small{\textsf{\begin{pspicture}(0,0)(10,4.5)
   \rput(0,1){\rnode{a}{\psdblframebox[linewidth=1pt]{\parbox{2cm}{\centering{Statistical\\ Learning\\ Rule}}}}}
\rput(3.5,1){\rnode{b}{\psshadowbox[linewidth=2pt,framearc=0.3,fillstyle=solid,fillcolor=lightgray,framesep=0.2]{Knowledge}}}
   \rput(7,1){\rnode{c}{\psdblframebox[linewidth=1pt]{\parbox{2cm}{\centering{Prejudiced\\ Learning\\ Rule}}}}}
\rput(10,1){\rnode{d}{\psshadowbox[linewidth=2pt,framearc=0.3,fillstyle=solid,fillcolor=lightgray,framesep=0.23]{Belief}}}
   \rput(10,3.5){\rnode{e}{\psdblframebox[linewidth=1pt]{\parbox{2cm}{\centering{Decision\\ 
Rule}}}}}
   \rput(3.5,3.5){\ovalnode[linewidth=1pt]{f}{\parbox{2cm}{\centering{World\\ State}}}}  
   \rput(7,3.5){\rnode{g}{\psframebox[linewidth=1pt,framearc=0.15,framesep=0.2]{Action}}}
   \rput(0,3.5){\rnode{h}{\psframebox[linewidth=1pt,framearc=0.15,framesep=0.2]{Reaction}}}
   \ncline{->}{a}{b}
   \ncline{->}{b}{c}
   \ncline{->}{c}{d}
   \ncline{->}{d}{e}
   \ncline{->}{e}{g}
   \ncline{->}{g}{f}
   \ncline{->}{f}{h}
   \ncline{->}{h}{a}
  \end{pspicture}}}
\caption{Prejudiced learning in the context of decision making under uncertainty. 
Shaded rectangles are
internal states of the agent, unshaded ones stand for actual events.
Framed rectangles are algorithms.}
\label{fig:pred}
\end{figure}
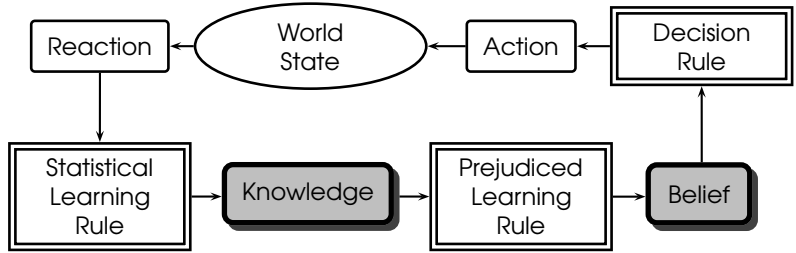

Given these conflicting goals above, it can be sensible to 
use what we would like to call 
\textbf{prejudiced learning system} for weights $\overline{w}_k$, 
by which we mean the following. 
The agent is given a start value $w_k(0)$ for its \textbf{belief} about
the value of $\overline{w}_k$. 
At time $t$ it infers its (empirical) \textbf{knowledge} 
$\eta_k(t)$ from the $t+1$ observations made up to this time using some statistical 
learning rule not further specified, with the single requirement
that $\eta_k(t)$ asymptotically approaches the real value $\overline{w}_k$
as $t\to\infty$.
The agents then updates its \textbf{belief} $w_k$
about $\overline{w}_k$ at time $t+1$ by \textbf{prejudiced learning rule}
\[
w_k(t+1)=L\bigl((w_k(s))_{s\leq t},\eta_k(t)\bigr),\quad t=0,1,\ldots
\]
The essence of this constructive definition of prejudice is the heuristically sound
distinction between knowledge and belief, close in spirit to~\cite{BON02}. 

Figure~\ref{fig:pred} shows a schematic view of the internal structure of the 
prejudiced agent and places it into the general context of decision making under
uncertainty.
The scope of our model below is merely the transition from knowledge to belief
through a learning rule of the above kind\ie the three components in the
lower right quadrant of the diagram. 
We will assume in particular that the reaction consists merely in informing 
the agent of its error\ie the numerical difference between its belief and the 
actual state of the world.
Furthermore, since we consider the parameters $\overline{w}_k$ to be independent from
and $w_k$ to be unrelated to each other, we focus on learning of a single
exterior parameter, and thus drop the index $k$ from now on.

Before proceeding, a caveat might be in order with respect to the terminology
we introduced.
The Webster's dictionary definition of the term ``prejudice'' is a 
``preconceived opinion, usually unfavourable; the holding of such an opinion;
an unjustified and unreasonable bias''.
But then, ``prejudiced learning'' seems self-contradictory from the outset,
and is certainly a controversial expression.
We nevertheless stick to this slight provocation, though the bare learning
rule above and the three axioms set out below would seem to only justify more 
cautious terms like ``prudence''.
We shall see that at least one very simple model for ``prejudiced learning''
will show characteristics, like staying away from the true value of the
parameter to be learnt, usually attributed to prejudice proper.
\subsection{Axioms}
The construction of the special model of a prejudiced learning rule 
we propose is based on three basic axioms and guided by the principle of simplicity.
The latter means in particular that we consider only \emph{memoryless}
prejudiced learning\ie $L=L(w(t),\eta(t))$. Furthermore, we seek to
render $L$ in the simplest (functional) form possible. 
The axioms for the model are derived from the heuristic meaning of 
three special cases of learning situations, cases which can be viewed as
constituting ancillary conditions under which a prejudiced rule has to 
function.
In particular, b) and c) are meant to justify the attribution ``learning rule'', 
while a) exhibits an extreme case of ``prejudice'':

a) \textbf{Inability is preserved:} The value $w(0)=0$ is taken to express 
the initial inability of the agent to perform the action. This must
remain constant\ie $w(0)=0$ implies $L(0,\eta(t))=0$.

b) \textbf{Importance spurs learning:} The higher an agents ranks
an action\ie the higher $w(t)$, the faster it shall adapt its belief
to its knowledge. That is, $\ABS{L(w(t),\eta(t))-w(t)}$ is
monotonously increasing in $w(t)$ for all $\eta(t)$.

c) \textbf{Truth is preserved:} If the agent's belief equals its 
knowledge, then it is kept constant\ie $L(\eta(t),\eta(t))=\eta(t)$.

These axioms imply a fundamental asymmetry  between high and low risk,
which will reemerge in the behavioural patterns of any model fulfilling
them, as will be seen in the particular case below. Heuristically, 
the axioms incorporate a certain cautiousness, in that they
tend to preserve a belief of high risk (low weight). Regarding axiom a),
note that we formulated it in the least restrictive way, since it
does not exclude a value $w(t)=0$ at later times. The first axiom also
entails a straightforward, implicit assumption on the decision rule, namely that
actions with weight zero are not taken.
\subsection{Noiseless Case: The Logistic Learning Map}
The combination of the axioms with the definition
of prejudiced learning is certainly satisfied by numerous functional
forms of learning rules. 
We now concentrate on a single special model which, because of its
simplicity allows to derive some phenomenological consequences which
seem to be of general importance and might pertain also to more complex
variants of prejudiced learning rules.
It should however be noted that any conclusion drawn from this special
model holds properly speaking only for this model and not for general
prejudiced learning rules, even if they satisfy the axioms above.

For the construction of a prejudiced learning rule satisfying
the above axioms, we specialise to the noiseless case\ie
we assume the input $\eta(t)=w=1-r$ to be identical to the true
weight for all times. Expressed in the believed risk $r(t)$ and 
its \textbf{error} $\Delta(t)\DEF r-r(t)$, the simplest prejudiced
learning rule is linear in either of these variables:
\[
\Delta(t+1)=\alpha\cdot r(t)\cdot\Delta(t),
\]
with a parameter $\alpha>0$. Re-expressed in the original
variables it reads
\begin{align*}
r(t+1)&=r-\alpha\cdot r(t)\cdot(r-r(t))\quad\text{or}\\
w(t+1)&=(1-\alpha)\cdot w+\alpha\cdot w(t)\cdot (1-w(t)+w).
\end{align*}
This rule reflects the cautiousness inherent in the axioms,
in tending to maintain a prejudice of high risk (low weight).
It satisfies conditions b) and c) and can be augmented by a simple
condition on $\alpha$ to also satisfy a), see \refS{viab} below.
Two extremal cases for the parameter $\alpha$ are $\alpha=0$,
in which case the agent immediately learns the input value $r$\ie
no prejudice is present, and  $\alpha=1/r(0)$ prohibiting any
learning. 

Introducing the \textbf{relative error} $\delta(t)\equiv\Delta(t)/r$ 
the learning map reduces to
\begin{align*}
\delta(t+1)&=\alpha\cdot\frac{\Delta(t)}{r}\biggl(r-\Delta(t)\biggr)=
\alpha r\cdot \frac{\Delta(t)}{r} \left(1-\frac{\Delta(t)}{r}\right)=\\
&=f_\rho(\delta(t))\equiv\rho\cdot\delta(t)(1-\delta(t)),\quad\text{with }\rho\equiv\alpha r.
\end{align*}
This is the well known logistic map~\cite[Section~7--4]{b:HEC90}, whose 
characteristics are entirely determined by the
bifurcation parameter $\rho$. It is usually considered as a self-mapping
of the unit interval, but note that the natural domain for $\delta$ is
here $[(r-1)/r,1]$.
\subsection{Classes of Behaviour}
\labelS{class}
We assume the intrinsic parameters $(\alpha,r(0))$ to be invariable
characteristics of a given agent. Then, in the noiseless case, the
behaviour of an agent is determined  by these intrinsic parameters, 
and the environmental parameter $r$ via $\rho=\alpha r$. We tentatively
distinguish between three classes of agents, and denote them heuristically
as follows.

\textbf{Adaptive (A):} For $\rho<1$, the logistic map has $0$ as a single,
attractive fixed point. These agents are therefore bound to adapt
to the external input $r$ at an exponential rate.

\textbf{Stubborn (S):} In the domain $1<\rho<3$, the fixed point $0$
becomes unstable, and the unique  stable fixed point is 
$\delta^\ast=(\rho-1)/\rho$, which is again approached at an exponential
rate. Agents of this class are bound to persistently underestimate
the true risk to a certain degree. 

\textbf{Uncertain (U):} For $4>\rho>3$, the logistic map exhibits
a bifurcating transition to deterministic chaos. Those agents
exhibit an increasingly erratic behaviour as $\rho$ rises, which
we subsume under the label `uncertainty'.

For $\rho>4$, the logistic map is no longer a self-mapping of the
unit interval. We simply ignore this case.

We will see in \refS{viab} that behaviours of the last two classes
 S and U can, under reasonable assumptions, only occur if the
agent initially underestimates the risk $r(0)<r$. On the other hand, 
an initial value $r(0)>r$ will lead to agents of class A, corroborating
the heuristics that cautiousness, that is overestimation of the risk,
entails a rather safe behaviour.
\subsection{Viability Conditions}
\labelS{viab}
In many senses, the logistic learning map is too simple
to work properly. In particular, it does not satisfy 
axiom a), but this can be accomplished by adapting the
intrinsic parameters $(\alpha,r(0))$. The value $r(0)=1$
implies $\alpha=1$. This condition, which we impose from now on,
is a paradigm for what we call a \textbf{viability condition}.
These are conditions on the intrinsic parameters
that guarantee a proper functioning of the
agents or improve their performance in a given environment.
A biological heuristics for their prevalence is that
agents not fulfilling them are naturally deselected.

We use two other viability conditions. First, we restrict
the range of admissible $\alpha$-values for all agent classes
to $\alpha<1/r(0)$, to make the learning map contractive at 
least in the first time step, and to omit the extremal case
of non-learning $\alpha=1/r(0)$ mentioned above.
Though we could restrict the range of $\alpha$ further to avoid
the occurrence of agents with $\rho>4$, we refrain from posing
the pertinent intrinsic condition $\alpha\leq 4$, since this
demand seems too restrictive for all classes.
As said above, we ignore agents with resulting
$\rho>4$.

The last condition concerns only
agents of class A and is somewhat more severe in its consequences:
For these agents,  the fixed point $\delta^\ast$ is negative and moreover
repulsive. If it lies within the range $[1-1/r,1]$ of admissible
$\delta$-values, then the learning map would diverge to 
$\delta\to-\infty$ whenever $r(t)$ becomes $<\delta^\ast$.
Although this could be avoided in the noiseless case by admitting only
start values $r(0)\geq\delta^\ast$, it would almost surely happen in 
noisy environments\eg the additive noise model of the
following subsection. 
To prevent this disastrous effect we require $\alpha<1$ for A-agents 
as the third viability condition.
This leads to $\delta^\ast<1-1/r$, pushing the repulsive fixed point
out of the range of $\delta$.

The viability conditions already have direct consequences for
S- and I-agents. Due to the condition $\alpha<1/r(0)$, the case
$\rho>1$ can only occur if $r/r(0)>\rho>1$\ie if the agent initially
underestimates the risk. Furthermore, The higher values of 
$\rho$, and therefore the more complex behaviour patterns, 
emerge with increasing discrepancy between initial belief
$r(0)$ and knowledge $\eta(t)=r$. We have given
a heuristic interpretation of these features in \refS{class}.

Although it would be desirable to dispense with the viability conditions
by refining the learning rule, this does not seem easy at the
given level of simplicity, without giving away other desired
features. For instance, consider the straightforward attempt
to adapt the parameter $\alpha$ with time to keep it
$<r(t)$\eg using the additional rule $\alpha(t+1)=\alpha(t)r(t)/r(t+1)$,
and thus force the learning map to be contractive
at all times. This would certainly enable us to lift the
viability condition $\alpha<1$ for A-agents.
Yet, apart from necessitating at least a one-step memory
for the prejudiced learner, it would also
let the modified agents become subject to fluctuations in
$r(t)$, something the heuristics for the construction of
prejudiced learning rules suggests to avoid in the first place.
We will propose a measure of the pertinent quality in
\refS{perf}, and see that the logistic learning map amended
by the viability conditions performs well with respect to it.

As opposed to these linear boundary conditions in parameter space,
we will employ boundary conditions in real space in the next subsection,
when submitting the system to a noisy environment.

We can now obtain a very coarse picture on the relative
proportions in which an observer would expect the three 
behavioural classes to occur in a population of prejudiced 
learners. We calculate the \emph{a priori} probabilities
for an agent to belong to one of the classes,
assuming that $(r,r(0),\rho)$ are uniformly distributed in
the domain $(0,1)^2×\{0\leq\rho<r/r(0)\}$, ignoring the
last viability condition that affects only A-agents. Integrating
over the pertinent ranges and discarding agents with a resulting
$\rho>4$ yields the ratios $4/5$, $8/45$, and $1/45$ for
class A, S, and I, respectively. Thus, this na\"ive estimation
renders adaptive behaviour prevalent, while stubborn and uncertain
behaviour occur with small but non-negligible probability.
\subsection{Adding Noise}
\labelS{noise}
We now assume that the input $\eta(t)$ of the prejudiced learning rule
underlies additive, statistical fluctuations around the true risk value
$r$\ie
\[
\eta(t)=r+\Xi(t),
\]
with a random variable $\Xi$. As the simplest possible noise model we
choose $\Xi$ to be symmetrically distributed with spread $\Sigma\geq0$ 
around $0$. That is, $\Xi$ is i.i.d.\ in $[-\Sigma,\Sigma]$. To keep
$\eta$ in $[0,1]$, this limits the range of admissible values for
$\Sigma$ to $0\leq\Sigma\leq\min(r,1-r)$.
Using relative coordinates $\delta(t)$ and $\xi(t)\equiv\Xi(t)\cdot\rho/r$, 
we can separate the fluctuations from the learning map:
\begin{align*}
\delta(t+1)=f_{\rho,\xi}(\delta(t))
&=f_\rho(\delta(t))+\frac{\xi(t)}{\rho}\cdot(\rho(1-\delta(t))-1)\\
&=f_\rho(\delta(t))+\xi(t)(\delta(t)-\delta^\ast).
\end{align*}
This is an example of a \textit{random dynamical system} (RDS). 
In this special case, it is a \textit{dynamical system with inhomogeneous noise}, 
for which the inhomogeneity $\delta(t)-\delta^\ast$ vanishes at the fix point. 
While systems with homogeneous, additive noise have been intensively studied, 
the inhomogeneous case is scarce in the literature.
The additive separation of the noise from the dynamical mapping means, in
particular, that we will still be able to make use of the tentative 
classification of \refS{class} for the noiseless case, to classify, 
at least partially, the behaviour of the agents under noise.

The change of variable from $\Xi$ to $\xi$ renders $\xi$ an i.i.d.\ random
variable with range $[-\sigma/2,\sigma/2]$, where $\sigma=2\rho/r\cdot\Sigma$.
Therefore, the general bounds $\alpha<1/r(0)$ and $\Sigma\leq\min(r,1-r)$ imply
$\sigma<2/r(0)\cdot\min(r,1-r)$ in general, and $\sigma<2\rho$ in dependence
of $\rho$. On the other hand, the \textbf{noise level} 
$\sigma$ is limited to $\sigma<2\min(r,1-r)$ for
A-agents by the viability condition $\alpha<1$.

Since the noise is in general non-vanishing at both boundaries of the
domain $[(r-1)/r,1]$ of $\delta$, the functions $f_{\rho,\xi}$ are
in general not self-mappings of this domain, and thus need to be
augmented by boundary conditions. For the performance analysis of 
A-agents in the next section we use boundary conditions of von Neumann type
\[
f_{\rho,\xi}((r-1)/r)=(r-1)/r, \quad 
f_{\rho,\xi}(1)=1,
\]
for all $\rho<1$, $\xi$. That means that when the system hits the boundary,
it remains there for one step and has, due to noise, a probability $\geq1/2$
to leave it in the next step.
\begin{figure}[tb]
\setlength{\unitlength}{1cm}
\begin{footnotesize}
\begin{picture}(14,10)
\put(0,10){\resizebox{21cm}{!}{\rotatebox{-90}{\includegraphics{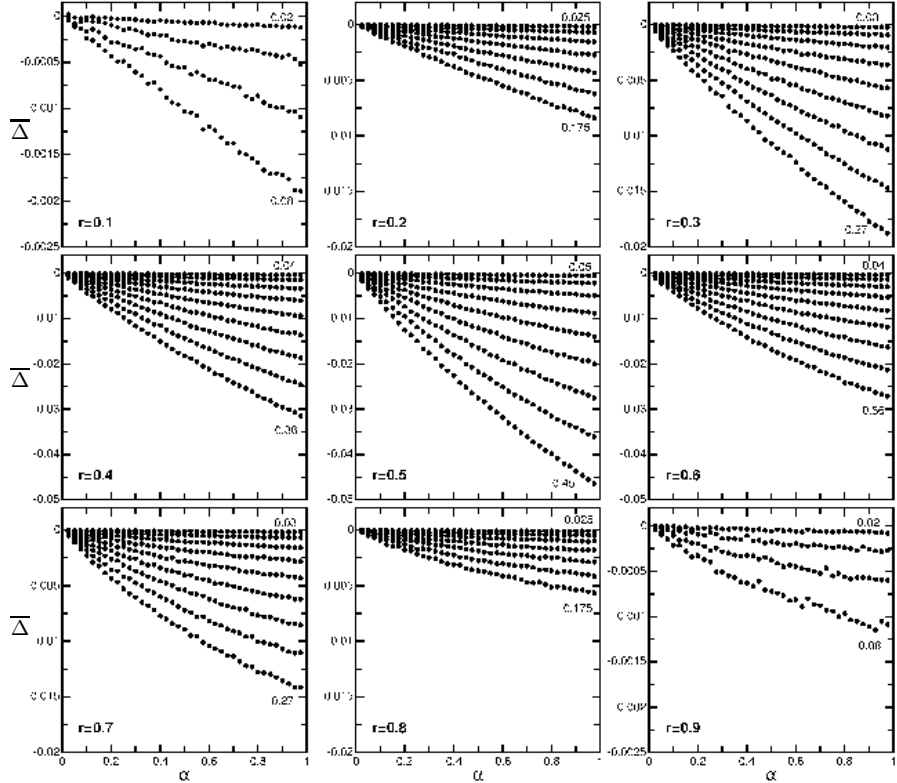}}}}
\put(0.2,7.9){$\overline{\Delta}$}
\put(0.2,4.6){$\overline{\Delta}$}
\put(0.2,1.3){$\overline{\Delta}$}
\end{picture}
\end{footnotesize}
\caption{Average error of A-agents. Numbers at
top/bottom curves denote minimum/maximum values
of $\Sigma$, between which the remaining curves 
interpolate in equidistant steps.}
\labelF{err}
\end{figure}
\section{Performance under Noise}
\labelS{perf}
\begin{figure}[tb]
\setlength{\unitlength}{1cm}
\begin{footnotesize}
\begin{picture}(14,10)
\put(0,10){\resizebox{21cm}{!}{\rotatebox{-90}{\includegraphics{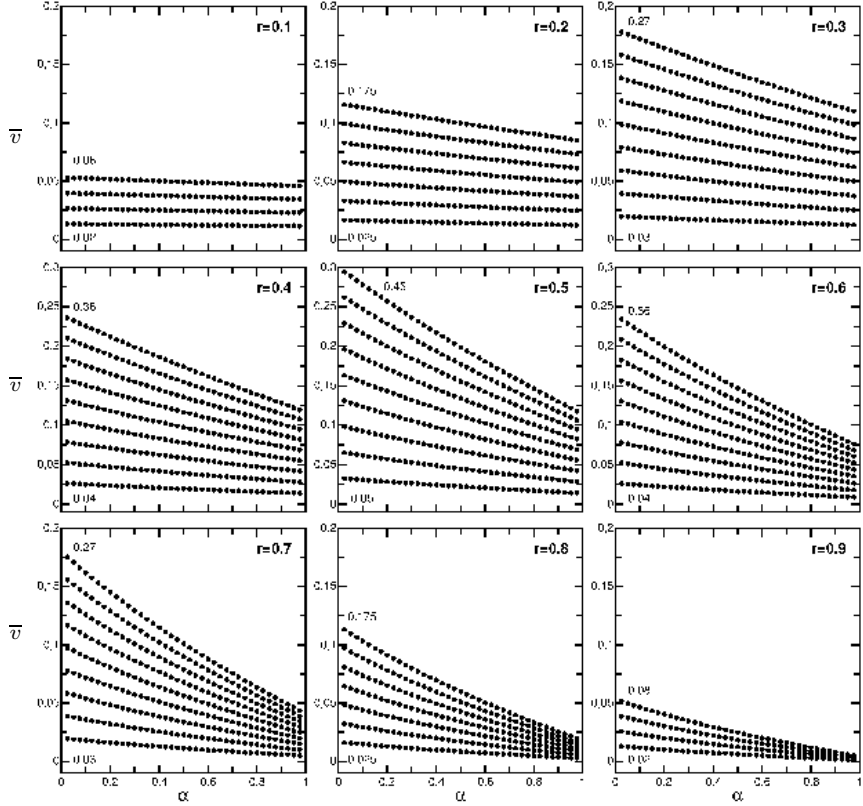}}}}
\put(0.2,7.9){$\overline{v}$}
\put(0.2,4.6){$\overline{v}$}
\put(0.2,1.3){$\overline{v}$}
\end{picture}
\end{footnotesize}
\caption{Average volatility of A-agents.}
\labelF{volat}
\end{figure}
Our prime heuristic for the introduction of the limitation of 
rationality presented by prejudice into learning was that agents
can benefit from not following stochastic fluctuations, in
addition to minimising the errors they make.
Therefore, the performance of prejudiced agents in a noisy environment 
should be assessed by considering at least the following \textit{two} 
natural quantities. 
The first one is simply the average error
$
\overline{\Delta}
$
which is a measure for the deterioration of learning performance
due to prejudice. However, we expect the learners to benefit from
prejudice by capitalising on a reduction of the average \textbf{volatility}
$
\overline{v}\DEF 
\overline{\ABS{r(t+1)-r(t)}}
$
of their belief, a variable which might for example 
be associated with an energy cost, perhaps arising from
an energetic price the agents would have to pay for
changing their selected action. 
We use these two variables to analyse the performance
of adaptive agents by numerical simulations.

The benign neglect of S- and I-agents is motivated 
by their fundamentally different behaviour that
hampers a quantitative comparison with A-agents.
In particular, stubborn agents have a constant error $\Delta>0$
and volatility $0$. This persists even when
they are subjected to noise, 
as will be seen, together with further qualitative features
of the other agent classes, in \refS{NDS}.

As a further restriction, following the heuristics that
statistical learning is slow in comparison
to the frequency of prejudiced learning and decision making, we consider the
limiting case in which the fluctuation level $\Sigma$ is constant,
and thus ignore statistical learning altogether.
Generically, we would suppose it to exhibit a slow decay, depending
on the efficiency of the statistical learning rule and the 
characteristics of the environment.

Simulations took place in the parameter ranges $0<r<1$,
$0<\alpha<1$, restricted by the viability conditions
of \refS{viab}, and $0<\Sigma<\min(r,1-r)$, determined 
by the noise model of \refS{noise}.
The results for error and volatility 
are shown in Figures~\ref{fig:err} and~\ref{fig:volat}, respectively,
in which every data point  represents an average over 
$2.5× 10^6$ time steps in 25 independent runs with
random starting values $r(0)$.

It springs to the eye in \refF{err} that the error induced by noise
is always negative\ie leads the agents to overestimate the risk.
 Furthermore, the error is always small, hardly ever reaching
$10$ percent, and tends to be somewhat smaller above $r=0.5$ than below.
Thus, the heuristics on cautiousness inherent in the construction
of the model is confirmed in this experimental
setting.

On the other hand, as the last two rows in \refF{volat} show, 
a stabilising mechanism of prejudiced learning becomes effective
with increasing $\alpha$, as $\overline{v}$ decreases from
its unadulterated value $2/3\Sigma$ at $\alpha=0$.
This is particularly true for higher $r$ entailing
a higher dynamical parameter $\rho$. In fact, as the graph
for $r=0.9$ exhibits, the learners become completely stable
when $\rho$ approaches $1$, even in the presence of noise. 
This remarkable feature will be studied further in \refS{NDS}.

Altogether, prejudice in learning has the potential to improve
the performance of agents by reducing their volatility significantly,
while not putting them far out in their risk estimation, or impeding
the efficacy of learning too much. In particular, 
the learning rate, 
or rather the rate at which a stable equilibrium is approached,
remains exponential.
\section{A Noisy Dynamical System}
\labelS{NDS}
\subsection{Noise Induced Stability of Prejudiced Learning}
\labelS{NIS}
In \refS{perf} we have seen that the prejudiced learning map
becomes increasingly stable as $\rho$ approaches $1$, despite
the presence of noise. In fact, the inhomogeneity of the 
noise, vanishing at the fixed point $\rho^\ast$, leads us
to the suspicion that this point plays a special role for
the dynamics of the map at higher $\rho$.
In this section we want to pursue this 
trail further, and consider the features of the map for 
$\rho$ between $1$ and $4$, the upper limit at which the noiseless 
logistic map reaches full deterministic chaos.

Here and in the
subsequent numerical analysis we employ periodic boundary conditions\ie
we consider the random RDS $f_{\rho,\sigma}$ defined by
\[
x_{t+1}=f_{\rho,\xi_t}(x_t)\DEF \rho x_t(1-x_t)+\xi_t(x_t-x^\ast)\mod 1,
\]
with $\xi$ an i.i.d.\ random variable in $[-\sigma/2,\sigma/2]$, and
$x^\ast=(\rho-1)/\rho$.
These boundary conditions make the presentation and the numerics somewhat 
simpler, while it turns out that they do not affect the behaviour of the 
system significantly. 

\begin{figure}[tb]
\setlength{\unitlength}{1.0cm}
\begin{picture}(12,10)
\put(-0.2,5.2){\resizebox{!}{5cm}{%
    \includegraphics{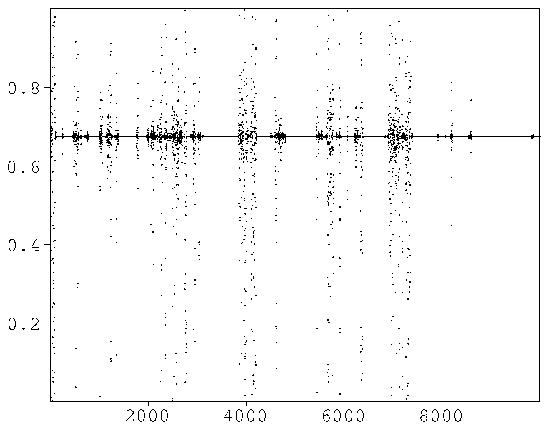}}}
\put(6.3,5.2){\resizebox{!}{5cm}{%
    \includegraphics{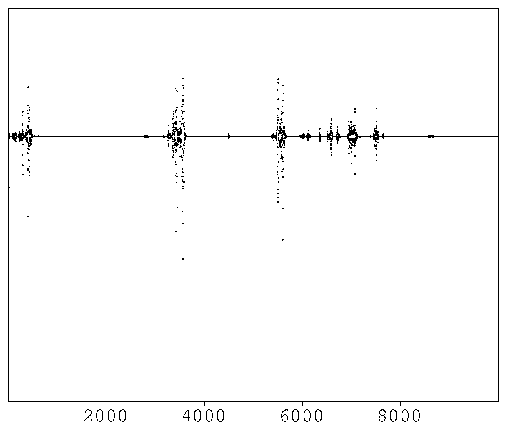}}}
\put(-0.2,0){\resizebox{!}{5cm}{%
    \includegraphics{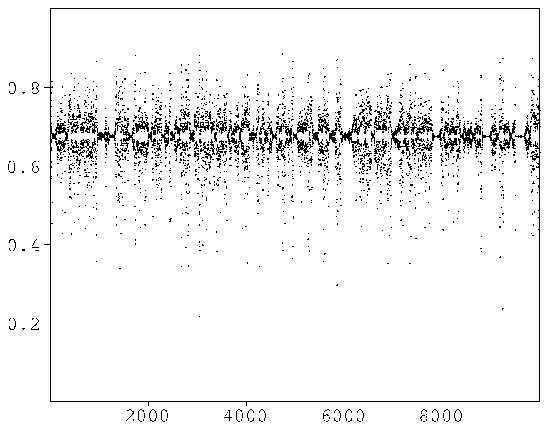}}}
\put(6.3,0){\resizebox{!}{5cm}{%
    \includegraphics{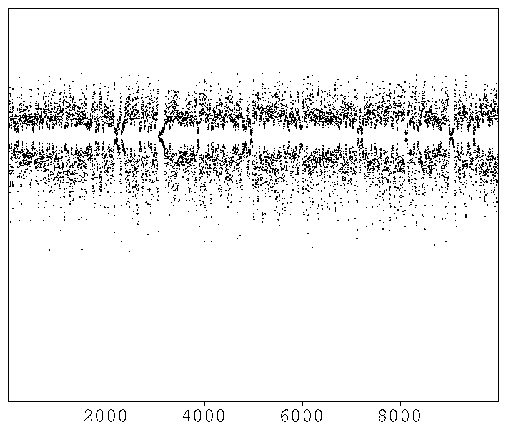}}}
\put(0.6,5.7){\makebox(0,0)[bl]{\small{a)\ \ $\sigma=4.8$}}}
\put(6.7,5.7){\makebox(0,0)[bl]{\small{b)\ \ $\sigma=1.37$}}}
\put(0.6,0.5){\makebox(0,0)[bl]{\small{c)\ \ $\sigma=1.1$}}}
\put(6.7,0.5){\makebox(0,0)[bl]{\small{d)\ \ $\sigma=0.7$}}}
\end{picture}
\caption{Examples for evolutions at $\rho=3.08$.}
\label{fig:exev}
\end{figure}

Simulations of this system show that, for many combinations of the dynamical
parameter $\rho$ and a nonzero noise level $\sigma$, it rapidly approaches,
and then stably remains at, the fixed point $x^\ast$. In fact, this holds for
\emph{all} admissible noise levels $\sigma<2\rho$ in the domain $1<\rho<3$
corresponding to agents of class S. Yet the effect prevails
even for a continuum of agents of class U, that is,
in the parameter range $3<\rho<4$, where the noiseless system undergoes
Hopf bifurcations into $2^n$ cycles on the route to deterministic
chaos. An example for this \textit{stochastic resonance} is shown in 
Figure~\ref{fig:exev}b), while Figures~\ref{fig:exev}c) and~d) show how
bifurcating behaviour is restored through a \textit{stochastic bifurcation},
when the noise level is lowered to leave the domain of the resonance
(the stability domain). These two phenomena are examined in the following
subsections.
\subsection{Basics of Noisy Dynamics}
\labelS{NDSBase}
Let us first introduce the necessary prerequisites on noisy, respectively,
random dynamical systems. We rely on~\cite{b:LM85} 
as a primary source, and develop the material using
the logistic learning map above as an example. Note that more refined
theoretical tools for the treatment of such systems abound by now,
see for instance~\cite{BBS02},
but are not needed in the analysis of the simple system at hand.

Every RDS, including the present, is completely characterised 
by its \textbf{transition density}
\begin{figure}[tb]
\setlength{\unitlength}{1cm}
\begin{picture}(12,6)
\put(-0.15,0){\resizebox{!}{6cm}{%
    \includegraphics{Transition_Density_Display_II.ps}}}
\put(6.15,0){\resizebox{!}{6cm}{%
    \includegraphics{Noise_Density_Display_II.ps}}}        
\put(10,0.5){\resizebox{!}{3cm}{%
    \includegraphics{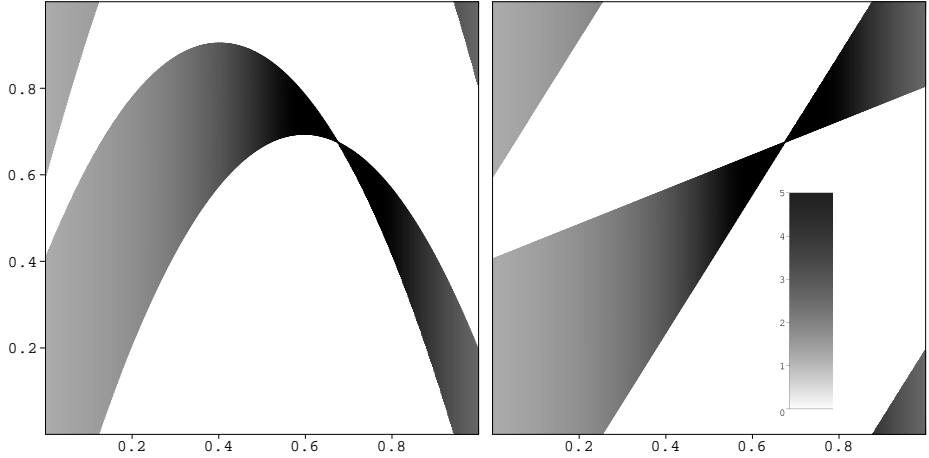}}}
\end{picture}
\caption{\PD[3.08,1.2]{} (left) 
and associated noise alone (right).}
\label{fig:Transition}
\end{figure}
\PD{x,y}, which yields the probability under $f_{\rho,\sigma}$
for ending up in an interval $J\subset[0,1]$ upon starting in 
$I\subset[0,1]$, by the formula 
\[
\Pr\nolimits_{\rho,\sigma}(I\to J)=\int_I\int_J \PD{x,y}\dd y\dd x.
\]
Figure~\ref{fig:Transition} shows an example of this density. 
Explicitly we find
\[
\PD{x,y}=(\sigma\ABS{x-x^\ast})^{-1}\cdot
\chi_{[x-\frac{\sigma}{2}\ABS{x-x^\ast},x+\frac{\sigma}{2}\ABS{x-x^\ast}]}(y),
\]
where $\chi$ is the characteristic function of an interval, and
periodic boundary conditions are implicitly assumed on all variables.
The transition density serves to define the 
\textbf{Perron--Frobenius (P--F) operator} of an RDS, 
a central tool in the system's analysis~\cite{CP83}. 
This operator acts on functions $u\in L^1([0,1])$\ie probability
densities by
\newcommand{\PF}{\ensuremath{\mathord{\textbf{PF}}}}
\[
\PF(u)(y)\DEF\int_0^1 \PD{f_\rho(x),y} u(x) \dd x. 
\]
The eigenvalues and eigenvectors of \PF\ are
of key importance. 
In particular, the positive and 
normalised eigenvectors to the highest eigenvalue $1$\ie 
probability densities $u$ with $\PF(u)=u$, 
are called \textbf{invariant densities} of the system. 
An invariant density $u$ defines an 
associated \textbf{invariant measure} 
$\mu_u$ by $\mu_u(A)\DEF\int_A u(x)\dd x$, where $A$ is
any Lebesgue-measurable set. An important example of
an invariant measure for the present system is $\mu_{\delta_{x^\ast}}$
generated by the the point measure $\delta_{x^\ast}$ at $x^\ast$,
as an easy calculation shows, using the fact that $\PD{x_n,\cdot}$
is a $\delta_{x^\ast}$-sequence if $x_n\to x^\ast$.
 
Yet more interest lies in the so-called \textbf{physically significant} 
or \textbf{Bowen--Ruelle--Sinai (BRS) measures}. A BRS-measure 
$\mu_{\text{BRS}}$ is defined 
for an ordinary\ie non-random dynamical system, 
defined by a deterministic mapping $f$,
by the following property. 
There exists a subset $U$ of the configuration space considered and with positive 
Lebesgue measure, such that for every continuous function $\psi$,
the following holds
\[
 \lim_{N\to\infty}\frac{1}{N}\sum_{j=0}^{N-1}\Psi(f^j(x))=
\int\Psi\dd\mu_{\text{BRS}}, 
\]
for all starting points $x\in U$, where $f^j$ denotes the $j$th iterate of 
$f$, see~\cite{DJ99}. By the Birkhoff individual ergodic theorem, this property is always 
fulfilled for $\mu_{\text{BRS}}$\emph{-almost all} $x$. 
The crucial strengthening of the hypothesis lies in 
the assumption that the ergodic hypothesis can be safely applied for \emph{all} 
starting points $x$ in a set of positive Lebesgue measure. 
For a RDS, the above property must be 
formulated in the mean with respect to the stochastic perturbation\ie the noise. 
Since most physically interesting quantities of a dynamical system are time averages, 
an ergodic hypothesis is regularly invoked by physicists to calculate them by space 
averages. This makes the existence and uniqueness of BRS-measures an important theoretical
issue in the study of random and ordinary dynamical systems. 
%
\subsection{Stable Phase and Lyapunov Exponent}
\labelS{Lyapunov}
\begin{figure}[tb]
\setlength{\unitlength}{1.0cm}
\begin{picture}(12,6)
\put(0,0){\resizebox{7cm}{6cm}{%
    \includegraphics{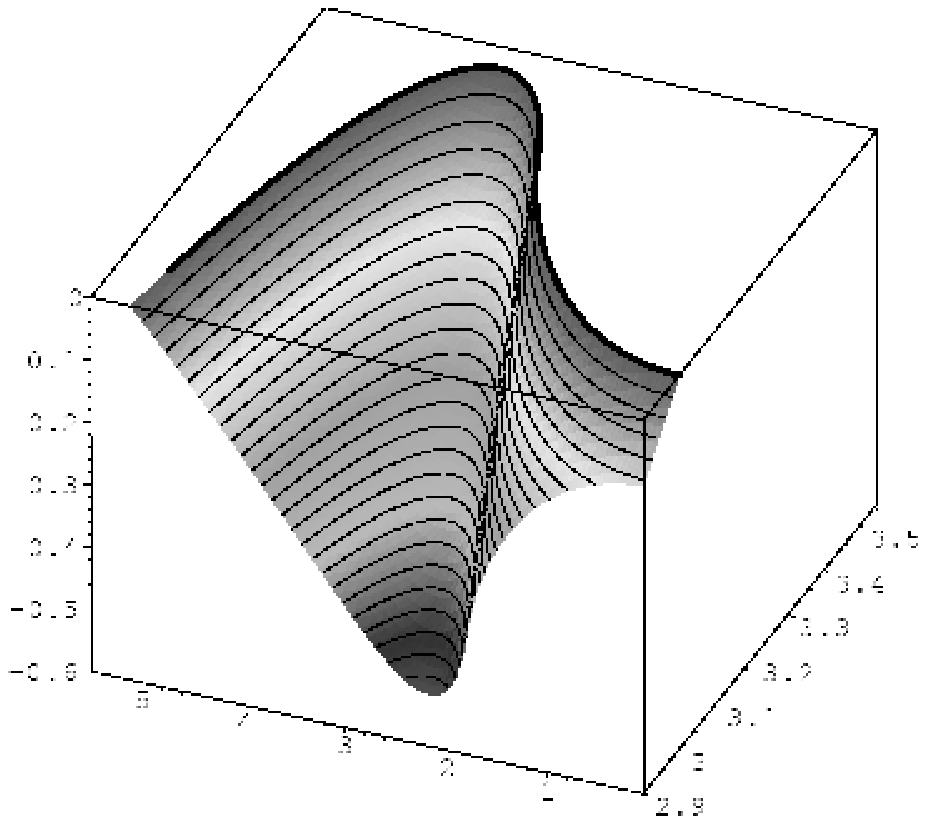}}}
\put(7.1,0.5){\resizebox{5cm}{5cm}{%
    \includegraphics{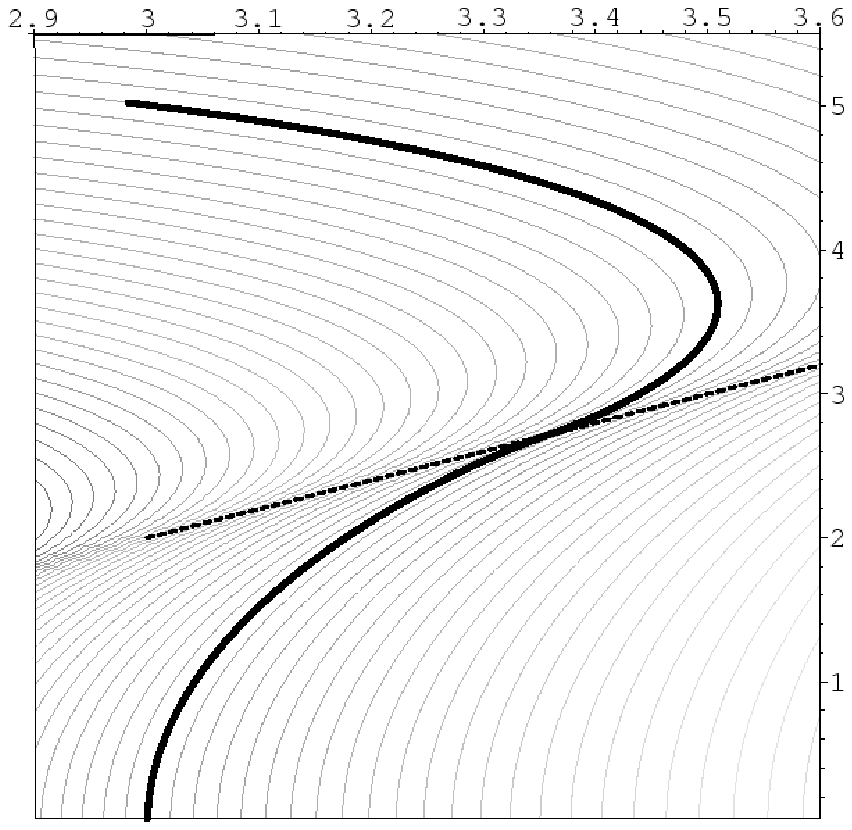}}}
\put(9.65,5.6){\makebox(0,0)[b]{\footnotesize{$\rho$}}}
\put(12.25,3.0){\makebox(0,0)[l]{\footnotesize{$\sigma$}}} 
\put(6.25,0.9){\makebox(0,0)[l]{\footnotesize{$\rho$}}}
\put(2.9,0.3){\makebox(0,0)[t]{\footnotesize{$\sigma$}}}
\put(-0.05,2.5){\makebox(0,0)[r]{\footnotesize{$\lambda_{\text{S}}$}}}          
\put(8.8,3.3){\makebox(0,0)[cc]{\textbf{S}}}
\end{picture}
\caption{Stable Lyapunov exponent $\lambda_{\text{S}}$.}
\label{fig:lambdaS}
\end{figure}
We are now in a position to bolster 
the heuristic conjecture of noise induced stability in the parameter range $3<\rho<4$
with theoretical and numerical evidence.
Furthermore, we want to determine the shape of the latter \textbf{stable phase}\ie
the set of $(\rho,\sigma)$ values for which the point mass concentrated at $x^\ast$ 
generates the unique BRS-measure, that is $\mu_{\text{BRS}}=\mu_{\delta_{x^\ast}}$.

The \textbf{Lyapunov exponent} $\lambda$ is central to the study of dynamical systems.
It is the measure for the exponential rate at which distant starting points
converge to an attractor under the system's evolution, respectively, the
rate at which nearby starting points become separated. In the former case,
$\lambda$ is negative, and in the latter positive, while zeros of $\lambda$ mark
transitions in the system's behaviour. Lyapunov exponents can be used to detect
stability, bifurcations, and the onset of chaos, and one of the simplest examples
for their use is, again, the logistic map~\cite[Section~7-4]{b:HEC90}.
It has been questioned whether Lyapunov exponents play the same role for
RDS, since there are examples in which a positive $\lambda$ does not indicate
ordinary chaotic behaviour~\cite{LPPV96,LPV96}. Yet, a \emph{negative} Lyapunov 
exponent always corresponds to stability, and therefore we chose to 
tentatively characterise the stable phase of our system through the
property $\lambda<0$. The quantity itself is defined by the
time average
\[
  \label{eq:lyapunov}
  \lambda\DEF\lim_{N\to\infty} 
  \frac{1}{N}\sum_{t=0}^{N-1}\ln\ABS{\smash{f_{\rho,\xi_t}'(x_t)}}.
\]
If a BRS-measure is known, $\lambda$ can be calculated, using an ergodic
theorem, as a space average
\begin{displaymath}
  \lambda=\left\langle
            \int_0^1 \ln\ABS{\smash{f_{\rho,\xi}'(x)}}\dd \mu_{\text{BRS}}(x)
          \right\rangle_\xi,
\end{displaymath}
taking into account that in the noisy case we also have to average over the 
random variable $\xi$. 

\begin{figure}[tb]
\setlength{\unitlength}{1.0cm}
\begin{picture}(12,8)
\put(0,0){\includegraphics{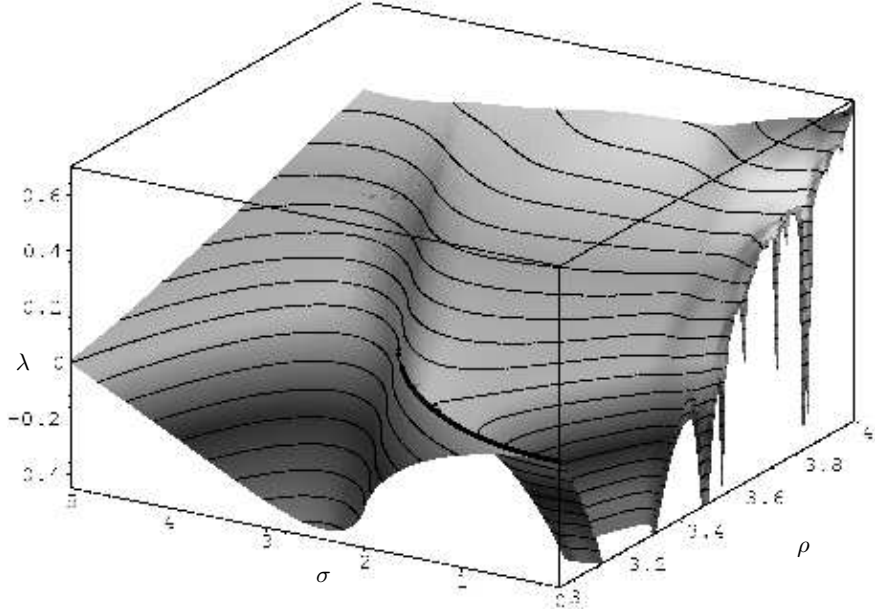}}
\put(4.3,0.5){\makebox(0,0)[tr]{\small{$\sigma$}}}
\put(0.3,3.3){\makebox(0,0)[r]{\small{$\lambda$}}}
\put(10.5,0.8){\makebox(0,0)[l]{\small{$\rho$}}}
\end{picture}
\caption{Numerical evaluation of $\lambda(\rho,\sigma)$. 
On a regular grid with resolution $(0.0025,0.05)$, 
each point represents $25$ independent runs of length $10^6$,after 
omitting $10^4$ initial iterations}
\label{fig:lambda}
\end{figure}
\newcommand{\ls}{\ensuremath{\mathord{\lambda_{\text{S}}}}\xspace}
The \textbf{stable Lyapunov exponent} $\lambda_\text{S}$ is now calculated
under the assumption $\mu_{\text{BRS}}=\mu_{\delta_{x^\ast}}$.
For the given noise model we obtain
\begin{align*}
  \ls(\rho,\sigma)&= 
    \frac{1}{\sigma}\int_{-\sigma/2}^{\sigma/2}
        \int_0^1 \ln\ABS{\smash{f_{\rho,\xi}'(x)}}\dd \mu_{\text{BRS}}(x)
    \dd \xi\\
&= \frac{1}{\sigma}\int_{-\sigma/2}^{\sigma/2}
        \ln\ABS{\smash{f_{\rho,\xi}'(x^\ast)}}
    \dd \xi \\
&= \frac{1}{\sigma}\int_{-\sigma/2}^{\sigma/2}
        \ln\ABS{\smash{\rho(1-2x^\ast)+\xi}}
    \dd \xi\\
&= \frac{1}{\sigma}\int_{-\sigma/2}^{\sigma/2}
        \ln(\rho-2+\xi)
    \dd \xi,\\
\intertext{for $\rho-2>\sigma/2$, and by a change of variable $\zeta=\rho-2+\xi$ 
we finally find, using the
abbreviations $\Delta_±=\rho-2±\sigma/2$}
\ls&= \int_{\ln\Delta_-}^{\ln\Delta_+}
        \frac{\zeta\ee^\zeta}{\sigma}\dd\zeta 
= \left[\frac{\zeta-1}{\sigma}\,\ee^\zeta\right]_{\ln\Delta_-}^{\ln\Delta_+}.
\end{align*}
Similar calculations in the two other cases $\rho-2<\sigma/2$ 
and $\rho-2=\sigma/2$ yield the net result
\begin{equation}
  \label{eq:lambdaS}
  \tag{$\ast$}
  \ls(\rho,\sigma)=\frac{1}{\sigma}
  \begin{cases}
    ( \ln\Delta_+ - 1 ) \ln\Delta_+ - ( \ln\Delta_- - 1 ) \ln\Delta_-,
      & \text{if $\Delta_->0$;}\\
    (\ln\sigma-1)\ln\sigma, & \text{if $\Delta_-=0$;}\\
    ( \ln\Delta_+ - 1 ) \ln\Delta_+ + ( \ln\ABS{\Delta_-} - 1 ) \ln\ABS{\Delta_-},
      & \text{if $\Delta_-<0$.} 
  \end{cases}
\end{equation}
\noindent Figure~\ref{fig:lambdaS} displays this result. 
The solid curve in both pictures is the nodeline $\lambda_{\text{S}}=0$
at which the hypothesis $\mu_{\text{BRS}}=\mu_{\delta_{x^\ast}}$
breaks down, while the dotted line in the right picture
is $\rho-2=\sigma/2$, where the two solutions in~\eqref{eq:lambdaS} connect.
The conjectured stability domain is marked by \textbf{S}.

We confirm the existence of a stable phase empirically by
collecting numerical data for the Lyapunov exponent using time
averages, obtaining the picture shown in Figure~\ref{fig:lambda}.
We find that $\lambda$ is identical to \ls within
the statistical error bounds for the most part of the region \textbf{S}.
Yet for higher $\sigma\gtrsim 3$, the empirical values of $\lambda$
are significantly larger than $\lambda_S$ along the inner boundary
of \textbf{S}. It seems plausible that the strongly intermittent behaviour of
the system in that region, of which Figure~\ref{fig:exev}a) shows
an example, together with the periodic boundary conditions, in
effect, prevents the convergence of $\lambda$ to $\lambda_S$.
\subsection{Stochastic Bifurcation}
\labelS{bif}
\begin{figure}[tb]
\setlength{\unitlength}{1.0cm}
\begin{picture}(10,8)
\put(-0.7,0){\resizebox{11cm}{!}{\includegraphics{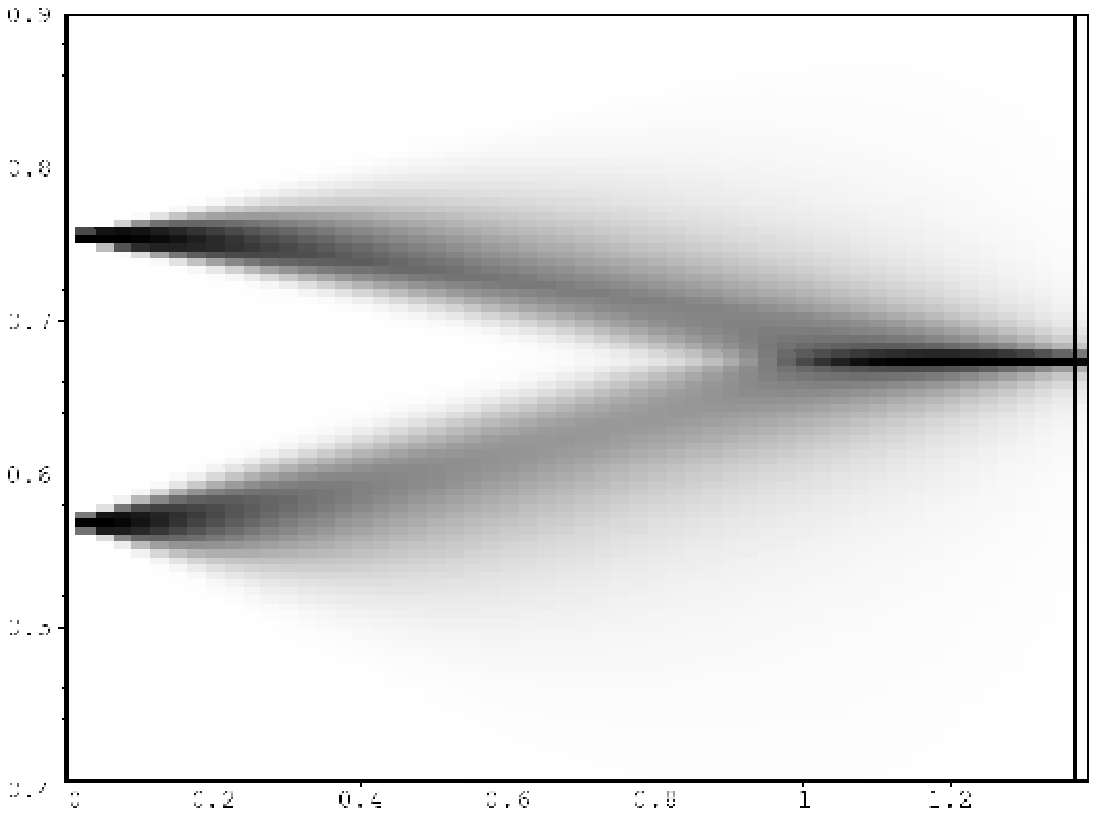}}}
\put(9.5,0.4){\resizebox{!}{3.5cm}{\includegraphics{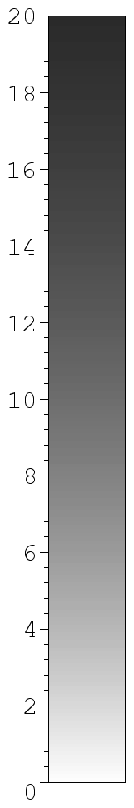}}}
\put(5,0){\makebox(0,0)[t]{\footnotesize{$\sigma$}}}
\put(10.1,0.15){\makebox(0,0)[t]{\small{$\sigma_0$}}}
\put(-0.4,4.3){\makebox(0,0)[r]{\small{$x$}}}
\end{picture}
\caption{Invariant densities at $\rho=3.08$
on a regular partition of $\{\sigma\in[0,1.4]\}×\{x\in[0.4,0.9]\}$
with resolution $(0.025,0.0025)$, for $N=1000$, and $M=4×10^6$.
Cells with absolute probabilities below $10^{-3}$ appear white.}
\label{fig:InvDens}
\end{figure}
The sample evolutions in Figure~\ref{fig:exev}b)--d) exhibit
the stochastic bifurcation the system undergoes when the
noise level is lowered to leave the stable phase. We now examine
this noise induced transition closer at $\rho=3.08$. It occurs at
the transition point $\sigma_0(\rho)$, which is obtained by numerically 
solving for the lower solution of $\lambda_S(\rho,\sigma_0(\rho))=0$ 
in~\refe{lambdaS}, yielding $\sigma_0(3.08)\approx 1.3683$.
As $\sigma$ approaches $0$, the system converges to the
deterministic $2$-cycle with attractor
$\bigl\{\rho_±=\frac{1}{2\rho}(\rho+1±\sqrt{(\rho+1)(\rho-3)})\bigr\}$.

\begin{figure}[tb]
\setlength{\unitlength}{2cm}
\begin{picture}(6,8)
\put(0,0){\resizebox{3.9cm}{!}{%
    \includegraphics{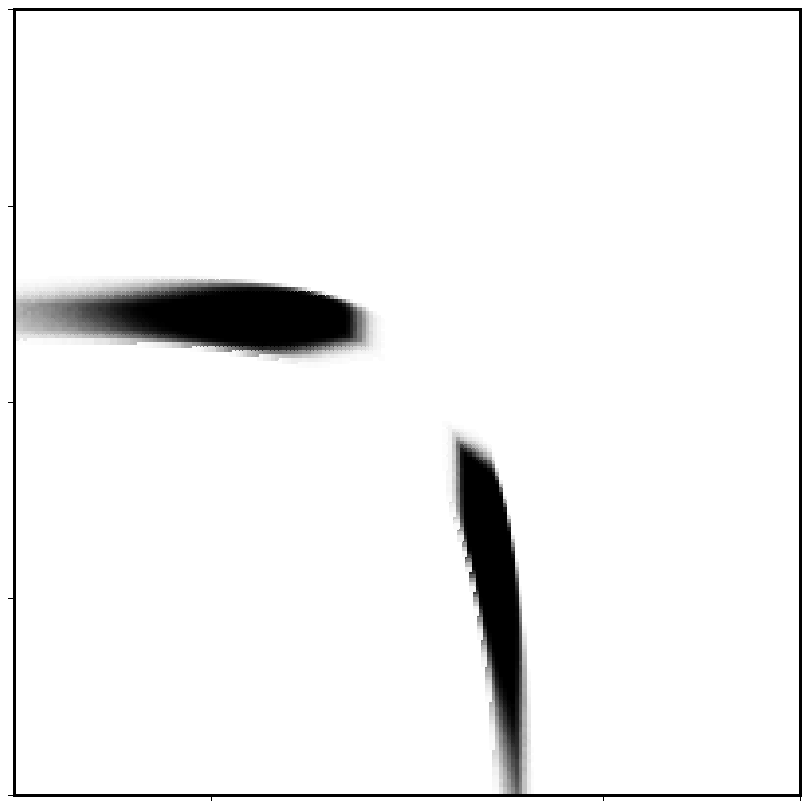}}}
\put(2,0){\resizebox{3.9cm}{!}{%
    \includegraphics{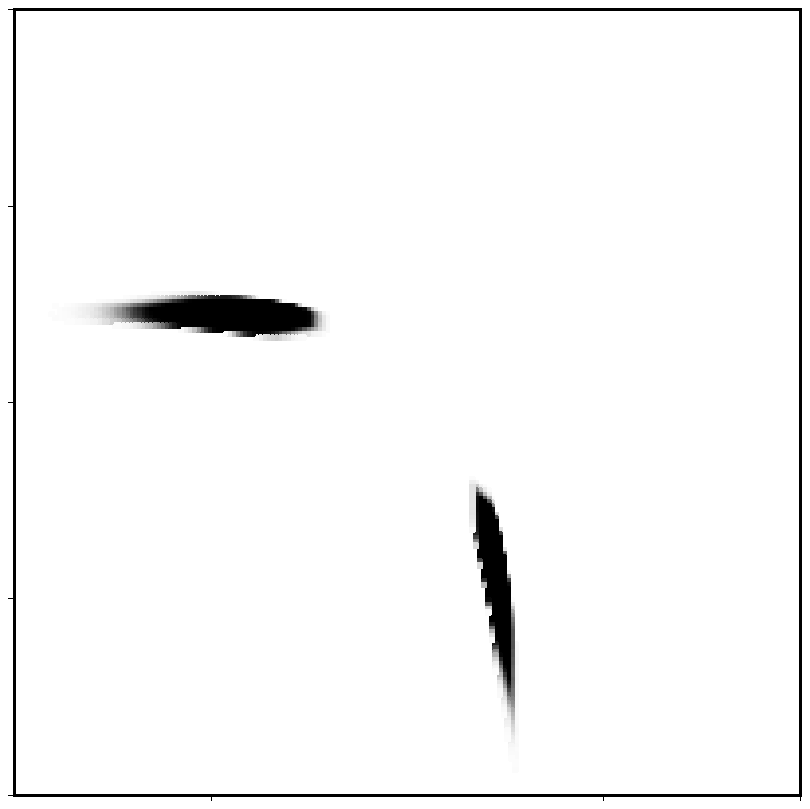}}}
\put(4,0){\resizebox{3.9cm}{!}{%
    \includegraphics{Return_Map_Density_Display_II1.ps}}}
\put(0,2){\resizebox{3.9cm}{!}{%
    \includegraphics{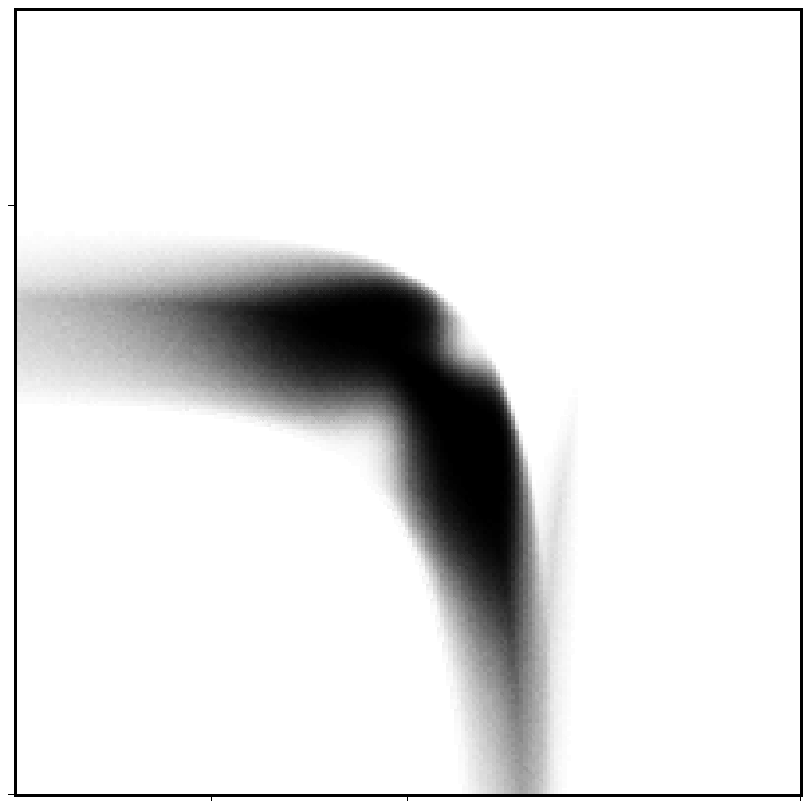}}}
\put(2,2){\resizebox{3.9cm}{!}{%
    \includegraphics{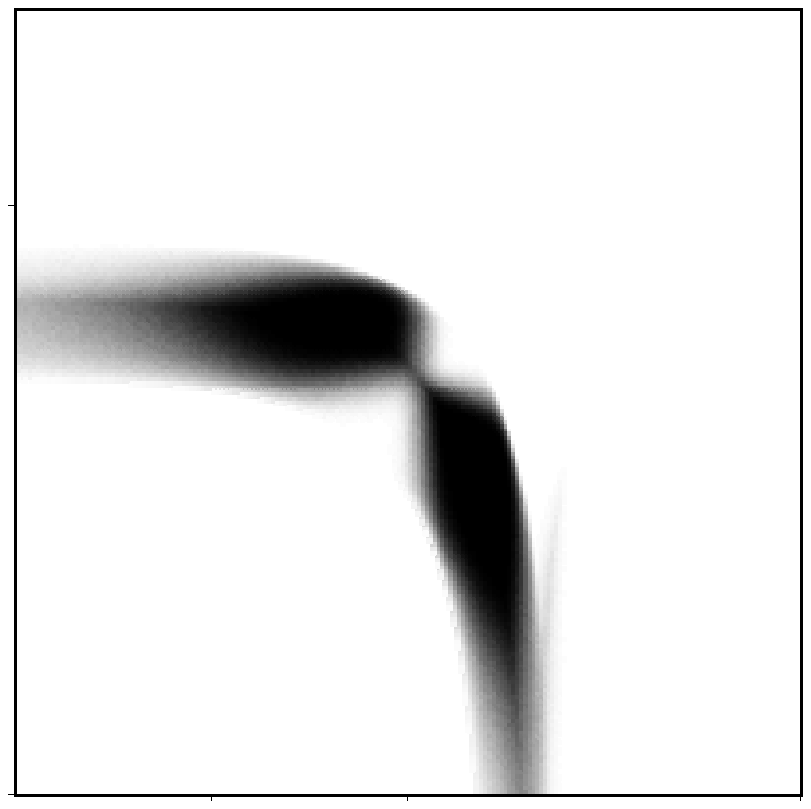}}}
\put(4,2){\resizebox{3.9cm}{!}{%
    \includegraphics{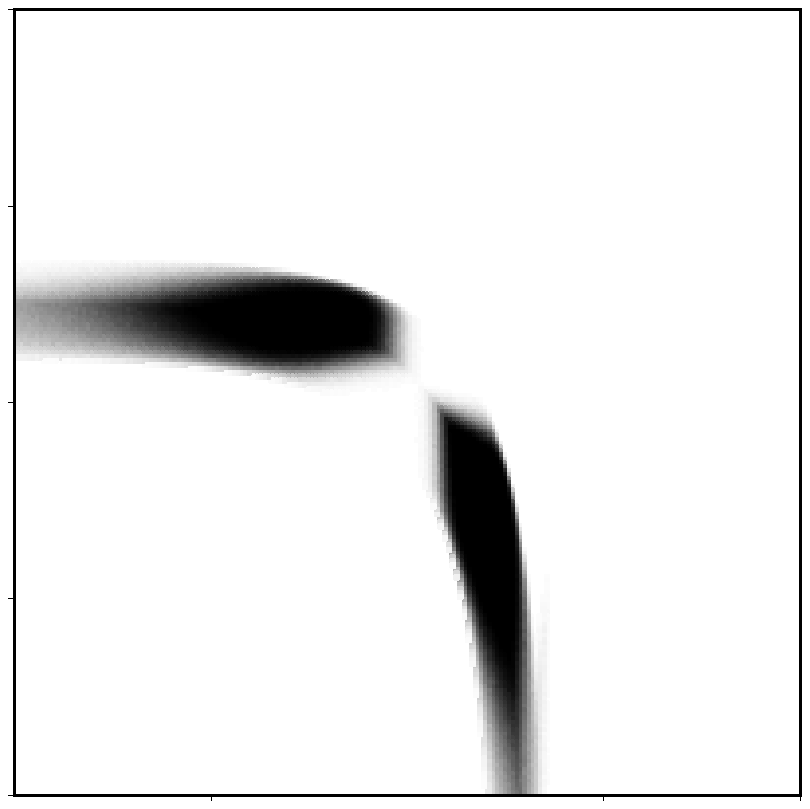}}}
\put(0,4){\resizebox{3.9cm}{!}{%
    \includegraphics{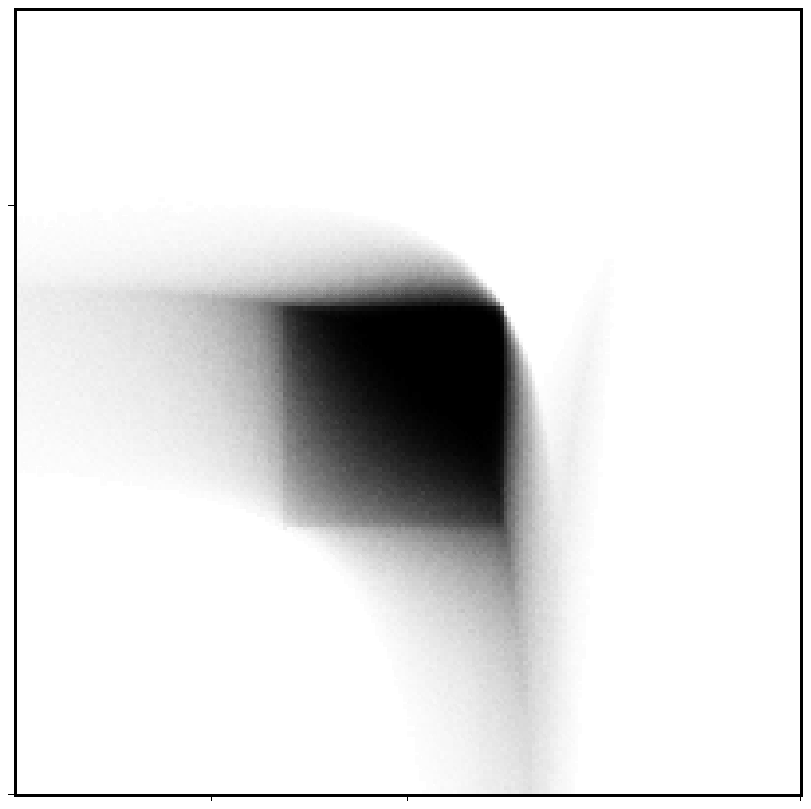}}}
\put(2,4){\resizebox{3.9cm}{!}{%
    \includegraphics{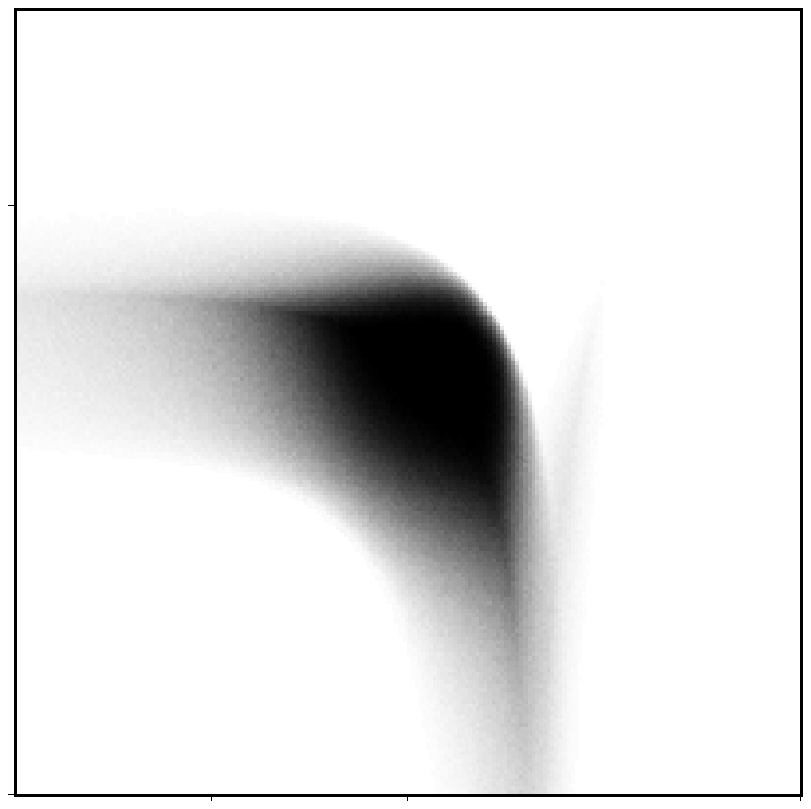}}}
\put(4,4){\resizebox{3.9cm}{!}{%
    \includegraphics{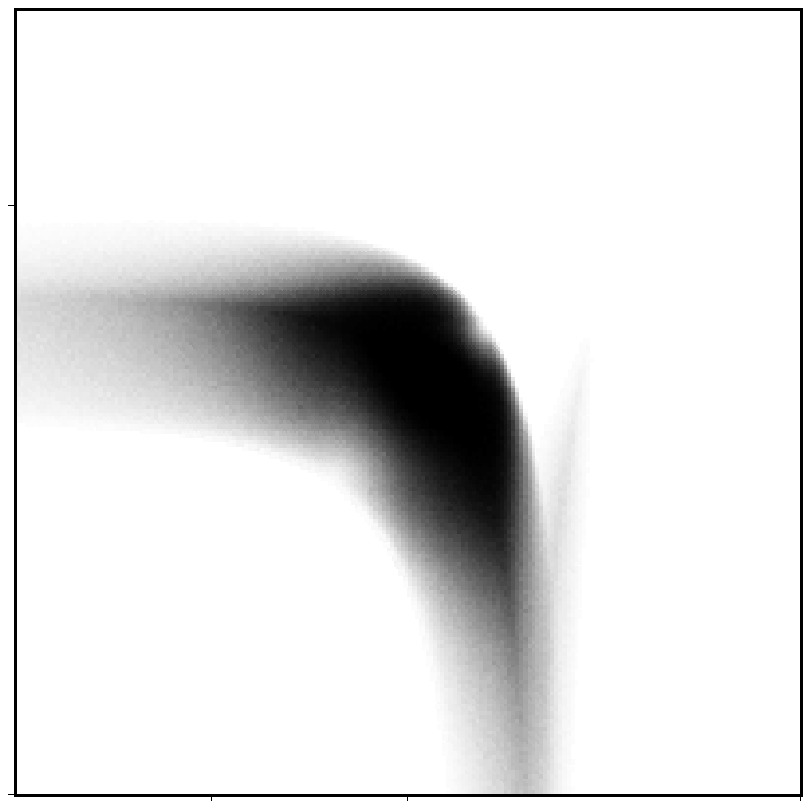}}}
\put(0,6){\resizebox{3.9cm}{!}{%
    \includegraphics{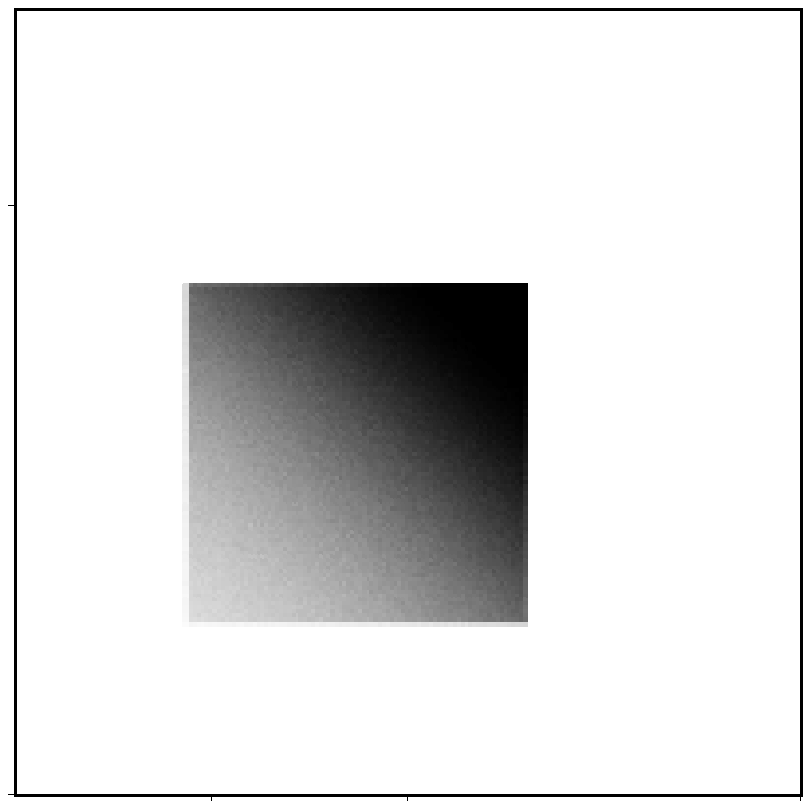}}}
\put(2,6){\resizebox{3.9cm}{!}{%
    \includegraphics{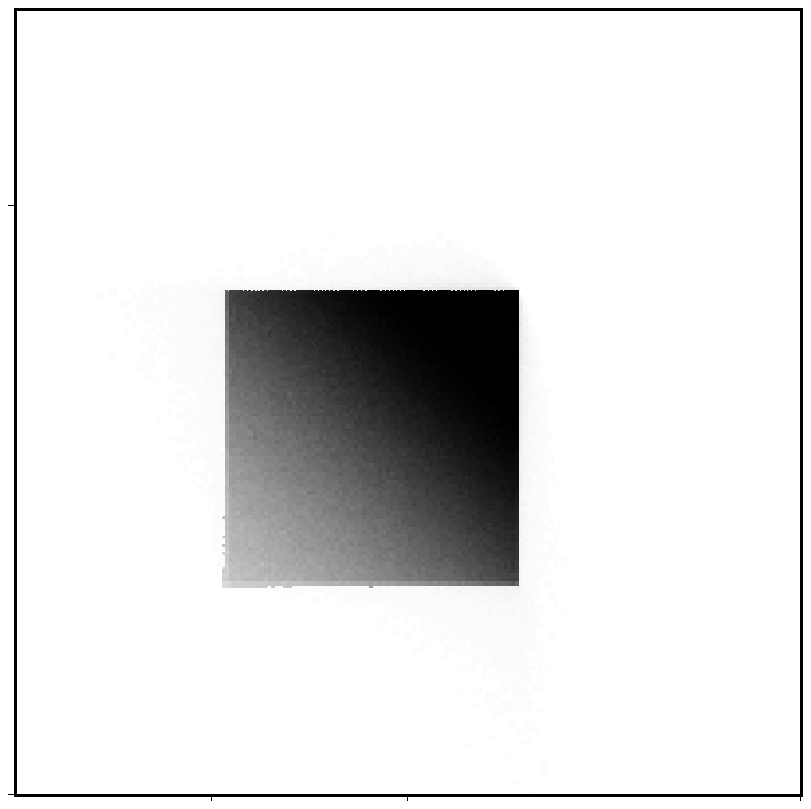}}}
\put(4,6){\resizebox{3.9cm}{!}{%
    \includegraphics{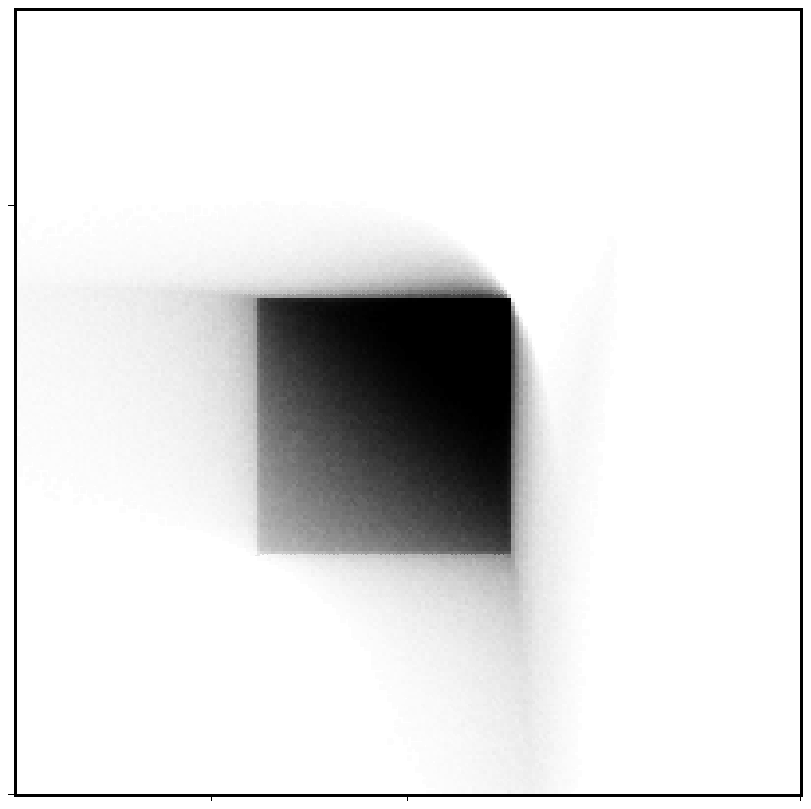}}}
\put(1.42,6.45){\resizebox{!}{2.8cm}{%
    \includegraphics{Return_Map_Density_Colorscale_II}}}
\put(1,-0.025){\makebox(0,0)[t]{\footnotesize{$y_{t}$}}}
\put(1.95,-0.025){\makebox(0,0)[t]{\footnotesize{$2$}}}
\put(-0.025,-0.025){\makebox(0,0)[tr]{\footnotesize{$-2$}}}
\put(-0.1,1.95){\makebox(0,0)[l]{\footnotesize{$2$}}}
\put(-0.1,1){\makebox(0,0)[l]{\rotatebox{90}{\footnotesize{$y_{t+1}$}}}}
\put(0.5,7.8){\makebox(0,0)[cc]{\footnotesize{$\sigma=1.5$}}}
\put(2.5,7.8){\makebox(0,0)[cc]{\footnotesize{$\sigma=1.\overline{36}$}}}            
\put(4.5,7.8){\makebox(0,0)[cc]{\footnotesize{$\sigma=1.2\overline{27}$}}}            
\put(0.5,5.8){\makebox(0,0)[cc]{\footnotesize{$\sigma=1.\overline{09}$}}}            
\put(2.5,5.8){\makebox(0,0)[cc]{\footnotesize{$\sigma=0.9\overline{54}$}}}            
\put(4.5,5.8){\makebox(0,0)[cc]{\footnotesize{$\sigma=0.\overline{81}$}}}            
\put(0.5,3.8){\makebox(0,0)[cc]{\footnotesize{$\sigma=0.681$}}}            
\put(2.5,3.8){\makebox(0,0)[cc]{\footnotesize{$\sigma=0.545$}}}            
\put(4.5,3.8){\makebox(0,0)[cc]{\footnotesize{$\sigma=0.4\overline{09}$}}}            
\put(0.5,1.8){\makebox(0,0)[cc]{\footnotesize{$\sigma=0.\overline{27}$}}}            
\put(2.5,1.8){\makebox(0,0)[cc]{\footnotesize{$\sigma=0.1\overline{36}$}}}            
\put(4.5,1.8){\makebox(0,0)[cc]{\footnotesize{$\sigma=0$}}}                        
\end{picture}
\caption{Transition densities of the return map at $\rho=3.08$,
approximated on a regular $200×200$ partition of $[-2,2]^2$ using 
$10×10^6$ iterations, omitting $10^4$ initial steps.
Cells with absolute probabilities below $10^{-6}$ appear white.}
\label{fig:Return_Map}
\end{figure}
Figure~\ref{fig:InvDens} shows the development of the invariant
densities for the BRS measures at $\rho=3.08$ as $\sigma$ varies.
The data for a given $\sigma$ represents the distribution of  $M$ 
i.i.d.\ starting points after evolving $N$ time steps. This 
statistical strategy can be justified if we assume that the 
Perron--Frobenius--Ruelle theorem~\cite{b:LM85} holds.
It asserts, among others, that the iterates $\PF^N(u)$ of any density 
$u$ of a normalised measure which is absolutely continuous
w.r.t.\ the BRS measure, converge to the density of the BRS measure
as $N\to\infty$. However, just above the transition point
(marked by the solid vertical line) the BRS measure $\mu_{\delta_{x^\ast}}$
is not well approximated with $N=1000$, cf.\ Figure~\ref{fig:exev}b).
Below $\sigma_0$, the singularity of the invariant density disappears,
and it develops two maxima around $\sigma\approx 1$, which become
separated at $\sigma\approx0.75$. Overall, the transition is a rather
typical example for a stochastic bifurcation~\cite[Chapter~9]{b:ARN98}.

Recently, it has been conjectured that stochastic bifurcations are in close
correspondence to phase transitions in systems of statistical 
mechanics~\cite{b:ZIN00}.
Although the theoretical basis for such a claim is presently not firm,
the two classes of  phenomena exhibit many common features. One of them
is \textit{symmetry breaking}, which is also present in the stochastic
bifurcations of our system. It can be exhibited by looking at the evolution
of the \textbf{return map} during the bifurcation. This map is defined by the 
sequence of the local expansion rates
\[
y_t\DEF \ln\ABS{f_{\rho,\xi_t}'(x_t)},
\]
and the statistical distribution of pairs $(y_t,y_{t+1})$ is its transition
density, for which some examples are shown in Figure~\ref{fig:Return_Map}.
The symmetry with respect to reflections on the diagonal that is present in
the stable phase, is broken by the stochastic bifurcation.
\subsection{Critical Exponent}
\labelS{crit}
\newcommand{\TP}{\emox{T_{\text{P}}}}
\newcommand{\DP}{\emox{D_{\text{P}}}}
\newcommand{\TPb}{\emox{\overline{T}_{\text{P}}}}
\newcommand{\DPb}{\emox{\overline{D}_{\text{P}}}}
\begin{figure}[tb]
\setlength{\unitlength}{1cm}
\begin{picture}(12,5.5)
\put(0.2,0){\resizebox{5.5cm}{!}{\includegraphics{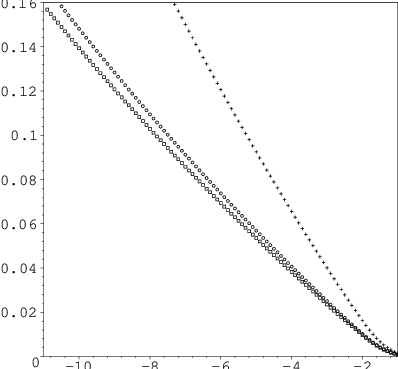}}}
\put(6.2,0){\resizebox{5.5cm}{!}{\includegraphics{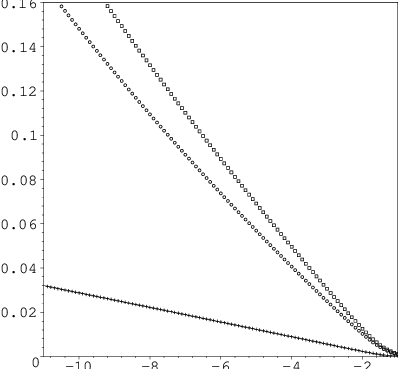}}}
\put(-0.2,2.2){\rotatebox{90}{\small{$\log_{10} \TP$}}}
\put(3.4,-0.1){\makebox(0,0)[ct]{\small{$\log_{10} r$}}}
\put(9.4,-0.1){\makebox(0,0)[ct]{\small{$\log_{10} r$}}}
\end{picture}
\caption{Left: Recurrence time statistics at $\tau=0$; $N=5×10^3$, $m=5×10^4$ ($+$); 
 $N=10^5$, $m=5×10^4$ ($\circ$);
 $N=5×10^6$, $m=10^3$ ($\Box$). 
Right: Correction for $N=10^5$, $m=5× 10^4$;
$\tau\approx0.00291$ ($\Box$); $\tau=0$ ($\circ$); \TPb ($+$) and linear fit.}
\label{fig:Recurrence}
\end{figure}
Finally, we want to further stress the analogy between stochastic 
bifurcations and critical phenomena in statistical mechanics~\cite{b:BDFN92}.
A characteristic of the latter is the vanishing or divergence of certain
intensive observable quantities, called \textit{order parameters}, at the 
critical point.  Their values near the critical point are governed 
by universal scaling laws, which are 
quantitatively described by the so called \emph{critical exponents}.

As a very simple example demonstrating the general scheme, 
consider the first bifurcation of the 
noiseless logistic map at $\rho_c=3$. The only relevant order parameter
is the Lyapunov exponent, which vanishes at $\rho_c$ with a critical
exponent
\[
\gamma_±\DEF\lim_{\tau\to 0_±} \dfrac{\ln \lambda(\tau)}{\ln \ABS{\tau}}=1,
\]
determined as the exponent of the leading power in the expansion of 
$\lambda$ in the scale-free parameter $\tau\DEF(\rho-\rho_c)/\rho_c$.

In our RDS, there are a number of other order parameters apart from $\lambda$ that
can be considered, and we use the \textit{Poincar\'{e} dimension} of the 
critical point $x^\ast$, which is a very natural measure for the collapse
of the invariant density to $\delta_{x^\ast}$ at the critical point.
It is determined by the \textit{recurrence time statistics} 
as follows, cf.~\cite{GAO99}.
The \textbf{Pioncarè recurrence time}, which is the average
time after which the system returns to a small domain 
$B_r(x^\ast)\DEF\{x\mid\ABS{x-x^\ast}\leq r\}$
of radius $r$ around $x^\ast$, is given by
\[
\TP(r)\DEF\lim_{N\to\infty}\dfrac{N}{\# \{x_i\in B_r(x^\ast),\ 1\leq i\leq N\}}.
\]
Obviously, this is not a scale-free quantity, 
so to obtain the desired order parameter, 
one assumes that \TP behaves asymptotically as
\[
  \TP(r)\propto r^{-\DP}\quad (r\to0).
\]
The number \DP so defined is the 
\textbf{Poincaré dimension} of $x^\ast$.
As the system enters the stable phase, 
with scale-free parameter 
$\tau\DEF(\sigma_0(\rho)-\sigma)/\sigma_0(\rho)$ 
tending to $0$ from above, 
$\TP(r)$ approaches $1$ for all $r$, 
and we expect \DP to vanish.

\begin{figure}[tb]
\psset{xunit=1cm,yunit=1cm}
\begin{pspicture}(0,0)(12,8)
\put(0,0){\resizebox{12cm}{!}{\includegraphics{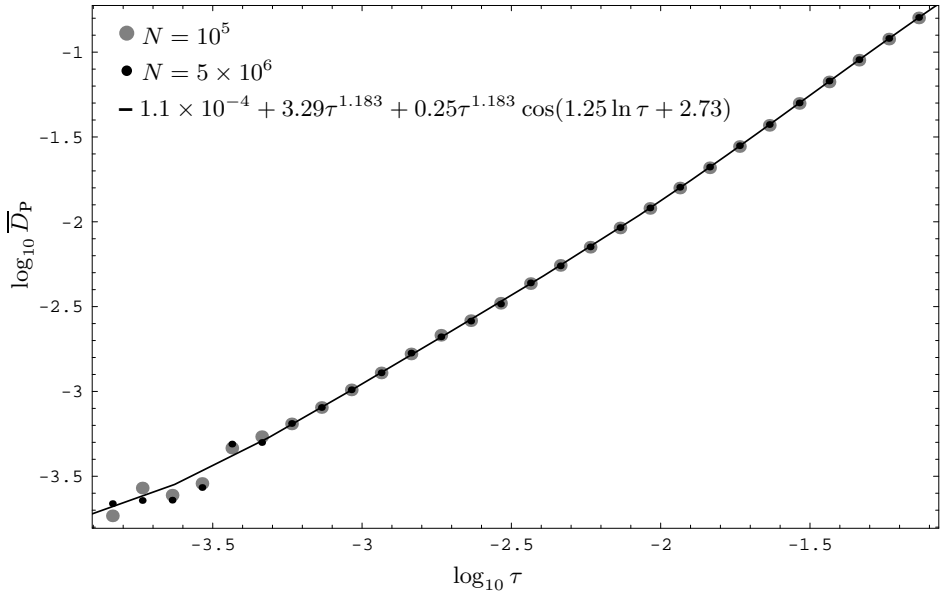}}}
\put(-0.4,3.6){\rotatebox{90}{\small{$\log_{10} \DPb$}}}
\put(6,-0.1){\makebox(0,0)[ct]{\small{$\log_{10} \tau$}}}
\newgray{halfgrey}{0.5}
\pscircle[fillstyle=solid,fillcolor=halfgrey,linecolor=halfgrey](1.2,7){1mm}
\pscircle[fillstyle=solid,fillcolor=black](1.2,6.5){0.7mm}
\put(1.4,7){\makebox(0,0)[lc]{\small{$N=10^5$}}}
\put(1.4,6.5){\makebox(0,0)[lc]{\small{$N=5×10^6$}}}
\put(1.1,6){\makebox(0,0)[lc]{\textbf{--}\ %
\small{$1.1×10^{-4}+3.29\tau^{1.183}+0.25\tau^{1.183}
\cos(1.25\ln\tau+2.73)$}}}
\end{pspicture}
\caption{Poincar\'{e}-dimensions and log-periodic fit at $\rho=3.08$.}
\label{fig:logper_fit_308}
\end{figure}
In the following, we examine the transition at $\rho=3.08$. Figure~\ref{fig:Recurrence}
shows various examples for recurrence time statistics gathered by numerical 
simulations with $m$ independent runs of length $N$, always omitting $10^4$ initial
steps. In the left picture we see three graphs for \TP at the critical point $\tau=0$
for various $N$. This exhibits the problem that, due to the combined effects
of slow convergence of the system as such, and the additional numerical error
which becomes relevant near the critical point, the measured Poincaré dimension
does not seem to converge to zero as $N\longrightarrow\infty$ (and does not
depend on the sample size for $m>10^4$). To circumvent this difficulty we tentatively 
replace \TP at given $\tau$ and $N$ with `corrected' values \TPb 
gained by division by \TP at $\tau=0$ and the same sampling time $N$, 
as shown in the right picture. A comparison of the `corrected' Poincaré dimensions
\DPb obtained from the \TPb for different values of $N$, 
see Figure~\ref{fig:logper_fit_308},
shows that these quantities do not depend strongly on $N$, 
lending some justification to this approach.

Remarkably it turns out that the behaviour of \DPb near $\tau=0$ is not governed
by a simple power law. Recently it has been observed in a number of areas that
singularities in many natural phenomena exhibit \textit{log-periodic oscillations}
corresponding to complex critical exponents, see~\cite[Section~3]{JS01} and
references therein. In such cases, one generally expects 
the considered observable 
to behave asymptotically like 
$\RE\tau^{\beta+\ii\omega}=\tau^\beta\cos(\omega\ln\tau)$.
Therefore, we fit the Ansatz
\[
\DPb(\tau)\propto A + B \tau^\beta + C \tau^\beta \cos(\omega\ln\tau+\phi)
\quad (\tau\to 0).
\]
to the dataset for \DPb. This yields the continuous line in 
Figure~\ref{fig:logper_fit_308} as the best fit graph, and
the corresponding value of the critical exponent is
\[
\beta = 1.18 ± 0.05.
\]
Notice that the modification of the pure power law is rather small,
as is expressed by the ratio $B/C\approx3.3/0.25=13.2$.
Similar analysis for $\rho=3.12$ confirms that
$\beta$ is within the given error bounds.
\section{Conclusions}
\labelS{Conc}
Reiterating that general conclusions about prejudiced learning rules cannot,
properly speaking, be derived by considering a single instance as we did above,
we still want to note some of the indications which the detailed study
of this special case provides us with.

The model for prejudiced learning and behaviour presented above
is rather plain by its construction, which was focused on
a few fundamental aspects we deemed characteristic.
Its utter simplicity, although of conceptual beauty,
appears as a drawback when its performance is critically
assessed. For instance, it made the introduction of 
viability conditions necessary to ensure the desired
functionality of the model, see \refS{viab}. These 
conditions are unsavoury in rendering the three 
different classes of prejudiced learners incomparable.
Yet, they are at least \textit{intrinsic} conditions
that can be satisfied by excluding certain values of
the model's parameters from consideration.
Totalling, the model itself calls for refinement to
be applicable, and performant, in more realistic situations.
Nevertheless, the basic performance results gathered in
\refS{perf} support the heuristics of model building
inasmuch as the class of adaptive agents learns at
exponential speed, and can benefit from prejudice in
a noisy environment by reducing their volatility 
significantly.

The most important phenomenological aspect of our model
is doubtlessly the possible emergence of stubbornness as a
\emph{secondary} phenomenon of prejudice. In fact,
for $\rho>1$, it becomes the prevalent behaviour in noisy
environments, aided by the mechanism of noise induced 
stability, a mechanism which is rather generic for nonlinear systems
driven by noise and thus could well apply for other
prejudiced learning rules than the logistic one.
It must be emphasised that
stubbornness cannot be ruled out as not being 
performant and therefore rare in realistic ensembles
of prejudiced learners. The only condition
for stubbornness to appear likely in individual agents
is that the risk is initially underestimated (by the pertinent
viability condition) and high, since $\rho=\alpha r$.
Stubbornness thus is a high risk phenomenon. 
One scenario particularly catches the imagination.
Assume for the moment that we have removed the awkward
viability condition $\alpha<1$ for adaptive agents\eg
by replacing it with suitable boundary conditions.
Then, agents of class A can be pushed
into classes S or U by a sudden rise of risk,
and consequently become stubborn.

The expectation to find the logistic prejudiced learning rule as such realised in
learning systems and environments as complex as human beings and 
human societies is doubtful. Yet it might still be a candidate model for
the description of certain behavioural aspects of biological systems
from the level of single-cells to that of plants and lower animals.
The thorough adherence to the principle of simplicity in the construction
of the model, realised in the rather reduced game-theoretic framework, the
minimal set of axioms, and finally the learning rule itself, in particular
its memorylessness, are in favour of that view.
Furthermore, the rich phenomenology of the model
can aid its identification in such systems through the provision of many 
indicators --- stubbornness being a prime one, alongside the characteristic
adaptive behaviour with attenuated volatility.

Let us conclude by indicating some directions for further research,
and posing some open questions.

In this first study, we have only considered static environments. 
One step to a more realistic model would be
to include the effect of a statistical learning rule in it.
This would generally lead to a slowly decreasing noise level,
and in turn could let agents undergo behavioural changes.
For example in class U, lowering the noise
leads to cyclic behaviour (`evasive') 
through stochastic bifurcations.

It would further be interesting,
along the lines sketched above, 
to improve the 
prejudiced learning rule itself by adding dynamical features.
Within the framework of our model, making the prejudice parameter
$\alpha$ dynamic naturally stands to reason,
although too simplistic approaches would be inappropriate,
cf.\ \refS{viab}. In view of the heuristics for the performance
of prejudiced learners developed in \refS{perf} it would seem promising,
for example, to lower $\alpha$ when the agent has lived through
a period of low volatility in the recent past. This modification
would in particular pertain to S-agents, whose behaviour would
then eventually become adaptive after some time in most
cases, unlike S-agents with fixed $\alpha$, and $\rho>3$, which normally enter the 
uncertain (bifurcating) mode when the noise level decreases.

While we have concentrated here on the behaviour and performance of
an individual prejudiced learner, studying the effects their presence
will have at the community level in a society of learning agents
is an important playground for further research. Good objects for such
a study would be small-world networks~\cite{WS98,NW99,NMW00,CS02} 
as models for propagation of information. A prime question here is 
to what extent a proportion of prejudiced learners can stabilise
beliefs in the given society. Studying these issues is work in progress.

This also, and finally, leads us to formulate some questions of an
evolutionary kind. What quantitative relations between agents
having various intrinsic parameters\ie
belonging to different classes, would emanate through
evolutionary pressure, over many generations of learners?
The viability conditions of \refS{viab} reflect our presumptions on the
net effect of evolutionary selection on the distribution
of $(\alpha,r(0))$, and for A-agents we have seen in \refS{perf}
that they have, in principle, a good chance to compete. Yet
both the former presumptions and the latter claim still have to
stand the test of more realistic models, including concrete
performance measures and evolutionary selection schemes.
%
%
\newcommand{\noopsort}[1]{} \newcommand{\singleletter}[1]{#1}
\providecommand{\bysame}{\leavevmode\hbox to3em{\hrulefill}\thinspace}
\providecommand{\MR}{Math.~Rev.\ }

\end{document}